\documentclass[a4paper,11pt]{article}
\usepackage{jcappub} 

\newcommand\lya{Ly$\alpha$}
\newcommand\nyx{\texttt{Nyx}}
\newcommand{\mpcph}{$h^{-1}$Mpc}
\newcommand{\kpcph}{$h^{-1}$kpc}
\newcommand\p[1]{$P_{\rm {#1}D}$}
\newcommand\px{$P_{\times}$}
\usepackage[normalem]{ulem}

\keywords{Lyman alpha forest, power spectrum}

\title{Measurement of the small-scale 3D Lyman-$\alpha$ forest power spectrum}

\author[a]{Marie Lynn Abdul Karim,}
\author[a]{Eric Armengaud,}
\author[a]{Guillaume Mention,}
\author[b]{Solène Chabanier,}
\author[c]{Corentin Ravoux}
\author[b]{and Zarija Luki{\'c}}

\affiliation[a]{IRFU, CEA, Université Paris-Saclay, F-91191 Gif-sur-Yvette, France}
\affiliation[b]{Lawrence Berkeley National Laboratory, 1 Cyclotron Road, Berkeley, CA 94720, U.S.A}
\affiliation[c]{Aix Marseille Universit{\'e}, CNRS/IN2P3, CPPM, Marseille, France}

\emailAdd{marie-lynn.abdulkarim@cea.fr}

\begin{document}
\abstract{Small-scale correlations measured in the Lyman-$\alpha$ (\lya) forest encode information about the intergalactic medium and the primordial matter power spectrum. In this article, we present and implement a simple method to measure the 3-dimensional power spectrum, \p3, of the \lya~forest at wavenumbers $k$ corresponding to small, $\sim$~Mpc scales. In order to estimate \p3 from sparsely and unevenly distributed data samples, we rely on averaging 1-dimensional Fourier Transforms, as previously carried out to estimate the 1-dimensional power spectrum of the \lya~forest, \p1. This methodology exhibits a very low computational cost. We confirm the validity of this approach through its application  to \nyx~cosmological hydrodynamical simulations. Subsequently, we apply our method to the eBOSS DR16 \lya~forest sample, providing as a proof of principle, a first \p3~measurement averaged over two redshift bins $z=2.2$ and $z=2.4$. This work highlights the potential for forthcoming \p3 measurements, from upcoming large spectroscopic surveys, to untangle degeneracies in the cosmological interpretation of \p1.}

\maketitle
\flushbottom

\newpage 

\section{Introduction}
\label{sec:intro}

The \lya~forest is a well-established observational tool in cosmology. When light from a distant object travels through a medium containing neutral hydrogen HI, it can undergo resonant absorption at a specific wavelength $\lambda_{\alpha} = 1215.67$~\AA. The HI component of the inhomogeneously-distributed intergalactic medium (IGM), located between a distant source and a terrestrial observer, therefore imprints its observed optical spectrum with the so-called \lya~forest~\cite{Gunn1965,Meiksin:2007rz}. Consequently, the spatial distribution of \lya~forest data is composed of sparse and irregularly distributed lines-of-sight throughout the IGM. This geometry complicates the determination of the \lya~forest 3-dimensional power spectrum, \p3. In this article, we implement a method to measure \p3 which addresses this issue by resorting to Fourier transforms, exclusively in one dimension along these lines-of-sight.

As a cosmological observable, the \lya~forest is currently the best probe of large scale structures, at least on a statistical ground, for $2< z \lesssim 4$. Within this specific redshift range, it is expected that the impacts of non-linearities and galaxy feedback on the matter distribution are less pronounced compared to $z\lesssim 1.5$, which is the main target of current galaxy surveys~\cite{BOSS:2016wmc}. Furthermore, hydrodynamical simulations have demonstrated that, except for the case of high-column density absorbers, the \lya~forest is produced mostly by the dilute part of the IGM, associated with under-dense or mildly over-dense regions of the cosmic web~\cite{Katz1995,Lukic2014}. This makes the \lya~forest sensitive to the matter power spectrum $P(k)$ at small scales $k\sim$~Mpc, in a way directly connected to the primordial linear matter power spectrum at these scales~\cite{Croft:1997jf,McDonald:1999dt}.

To study the small-scale matter distribution from \lya~forest observations, an efficient approach consists of measuring 1-dimensional correlations between absorption intensities within individual optical spectra of background sources. The statistical estimator of choice is the Fourier transform of the corresponding correlation function, the 1-dimensional power spectrum \p1$(k_{\parallel})$, where $k_{\parallel}$ is the wavenumber associated with the absorber's coordinate along a line-of-sight (LOS). Following the initial work by~\cite{Croft:1997jf,Croft:1998pe}, numerous \p1 measurements have been carried out over the past two decades, spanning a wide range of scales and redshifts.  Some of these measurements have focused on large $k_{\parallel}$, with relatively small samples of high-resolution, high signal-to-noise spectra, e.g.~\cite{McDonald:1999dt,Kim:2003qt,Walther:2017cir,Yeche:2017upn,Irsic:2017sop,Day:2019joh,Karacayli:2021jeg}. In contrast, 
some measurements were conducted down to smaller $k_{\parallel} \sim 0.1$~Mpc$^{-1}$, e.g.~\cite{SDSS:2004kjl,Palanque-Delabrouille:2013gaa,Chabanier:2018rga,Karacayli:2023afs,DESI:2023xwh}, relying on the large statistics achieved by cosmological spectroscopic surveys. Cosmological simulations including thermal properties of the IGM allow one to model the \lya~absorption field~\cite{Hernquist:1995uma,Hui:1997dp,Gnedin:1997td}. Since, as explained above, the measured \lya~forest comes from relatively dilute parts of the IGM, hydrodynamical simulations that do not incorporate small-scale galaxy physics and rely solely on an empirical photo-ionizing background are considered adequate for modeling \p1\cite{Lukic2014}, while the effects of galaxies and their feedback are expected to introduce $\sim 1-10$~\% level corrections~\cite{McDonald:2004xp,Chabanier:2020uuh}. The comparison of \p1 data with predictions from IGM simulations therefore permitted to infer the thermal properties of the IGM, e.g.~\cite{McQuinn2015,Walther:2018pnn}, while also constraining the cosmological primordial matter power spectrum. Many works relied on \p1~to either measure, in a relatively model-independent way, the amplitude and slope of the linear matter power spectrum at scales around $k\sim1$~Mpc$^{-1}$, e.g.~\cite{SDSS:2004aee,SDSS:2004kqt,Viel:2004bf,Pedersen:2022anu}, or constrain more specifically the properties of neutrinos~\cite{Viel:2010bn,Palanque-Delabrouille:2015pga,Palanque-Delabrouille:2019iyz} and dark matter~\cite{Viel:2013fqw,Irsic:2017ixq,Baur:2017stq,Armengaud:2017nkf,Irsic:2017yje,Murgia:2019duy,Rogers:2021byl,Hooper:2021rjc}.

However, \p1 only measures correlations between absorption features along individual lines-of-sight. Since the measured coordinates associated with the absorption field are wavelengths, and not physical positions, several physical effects are intertwined, leaving some ambiguity in the interpretation of the \p1 data. First, the wavelength separations between absorption features is the result of their velocity difference $\Delta v$, related both to physical separations in terms of comoving coordinates, and to velocity flows in the cosmic web associated with linear redshift-space distortions (RSD), and to non-linear velocity flows of the IGM. Second, the \lya~absorption field is smoothed on small scales along lines-of-sight, due to a blend of two very different physical effects: Jeans smoothing and thermal broadening. Jeans smoothing represents a physical smoothing of the IGM density field, and is due to the cumulative effect of thermal pressure over the cosmic history of the IGM. Thermal broadening is specific to spectroscopic observations, it is mostly dictated by the local temperature of the IGM. The variation of \p1 as a function of $k_{\parallel}$ is impacted by all these effects: disentangling them would improve the sensitivity and robustness of the above-mentioned cosmological constraints derived from \p1 measurements.

Given that correlations transverse to the lines-of-sight are differently affected by these effects, it is not surprising that the extension of \lya~correlation measurements to three dimensions was considered a long time ago~\cite{McDonald:1998pm,Hui:1998vf}. With the advent of the BOSS survey~\cite{BOSS:2012dmf}, a large sample of \lya~forests was available with a line-of-sight density of $\sim 15$~deg$^{-2}$. This enabled 3D correlation measurements on relatively large scales~\cite{Slosar2011}, for which the most relevant physical processes are RSD and the imprint of the baryon acoustic oscillations (BAO). First used for BAO measurements~\cite{Busca2013,Delubac2014,duMasdesBourboux:2020pck}, the \lya~forest 3-dimensional correlation function in real space, $\xi_{\rm 3D}(r_{\parallel}, r_{\perp})$, can also be exploited for RSD and the Alcock-Paczynski test~\cite{Gerardi:2022ncj,Cuceu:2022brl,Cuceu:2022wbd}. On the other hand, measuring 3D correlations on small scales, i.e.~for small angular separations, is more limited by statistics: although the absolute number of \lya~forest samples observed in modern surveys is impressive, there are few \textit{pairs} of \lya~forests with comoving transverse separation $r_{\perp} \lesssim 1\,{\rm Mpc}$. Even in a particularly dense SDSS field, called Stripe 82, the number density of measured $z>2.1$ quasar spectra is 37 deg$^{-2}$~\cite{Ravoux:2020bpg}: this results in a mean comoving transverse separation between nearest lines-of-sight of $8\,h^{-1}\,{\rm Mpc}$ at $z=2.2$. Line-of-sight pairs with significantly smaller angular separations are observed solely due to  fluctuations in the quasar sky distribution. However, the forthcoming spectroscopic surveys are expected to substantially increase the quasar density, for example the number density of $z>2.1$ quasars for the main DESI sample will be $n = 58$~deg$^{-2}$~\cite{DESI:2023dwi}. Given that the number of pairs scales like $n^2$, it is foreseeable that large-scale sky surveys will provide a sufficiently large statistical sample for the assessment of \lya~correlations at small $r_{\perp}$. Additionally, a small number of close pairs was observed in~\cite{Rollinde:2003tf,Becker:2004wk,Rorai:2017zxv}, with transverse separations $r_{\perp}$ in the $0.1-1$~Mpc range: the corresponding transverse correlations were used to estimate the Jeans scale.

A natural extension of \p1$(k_{\parallel})$ is the 3-dimensional power spectrum \p3$(k_{\parallel}, k_{\perp})$ of fluctuations in the \lya~absorption field. \p3 can be predicted from cosmological hydrodynamical simulations, similar to those used to model \p1~\cite{McDonald:2001fe,Givans:2022qgb,Chabanier:2022dgr}, but with different volume requirements. Several authors have then proposed analytical and physically motivated approximations~\cite{McDonald:2001fe,Arinyo-i-Prats:2015vqa,Garny:2020rom}, which turned out to fit well the results of full numerical simulations~\cite{Givans:2022qgb}. Therefore, there is a clear path for the interpretation of forthcoming \p3 measurements.
However, it is technically not straightforward to estimate \p3 from large-scale spectroscopic surveys. Apart from their large data volumes, \lya~forest samples possess an anisotropic geometry, involving a regular and dense sampling along individual lines-of-sight, which are themselves sparsely and randomly distributed over the sky. In~\cite{Font-Ribera:2017txs}, a method was proposed to compute \p3 from large samples. This method relies on initially estimating the so-called cross-spectrum \px, which is a hybrid quantity between \p3 and the real-space correlation function $\xi_{\rm 3D}$. In~\cite{Font-Ribera:2017txs}, the measurement of $P_{\times}$ is achieved through an optimal quadratic estimator, a technique that has already demonstrated success in the realm of \lya~forest observations for estimating \p1~\cite{SDSS:2004kjl,Karacayli:2021jeg,Karacayli:2023afs}. In this article, we introduce an alternative approach to estimate \p3. Similarly to the method proposed by~\cite{Font-Ribera:2017txs}, we begin by calculating \px~and subsequently derive \p3. However, our technique for estimating \px~relies on the 1-dimensional Fast Fourier Transform (FFT). Notably, the FFT method has previously been employed to compute \p1~from extensive samples of surveys like SDSS~\cite{Palanque-Delabrouille:2013gaa} and DESI~\cite{DESI:2023xwh}. The method presented here can thus be viewed as an extension to those prior works. In our case, even more importantly than in the case of \p1, one of the notable advantages of utilizing the 1-dimensional FFT approach is its minimal computational requirements, in addition to its inherent simplicity.

This article is laid out as follows: in Section~\ref{sec:method}, we present the basic definitions and formalism connecting \p1, \px~and \p3, and we describe the implementation of our method. Section~\ref{sec:NyxBoxValidation} demonstrates the performance of the algorithm on realistic hydrodynamical simulations, whose predicted \p3 is well-defined. In Section~\ref{sec:appli-SDSS}, we present an application of the method using \lya~spectra from the SDSS data, hence providing a first measurement of \p3 based on real data.

\section{Method}\label{sec:method}

The fundamental observable quantity for the \lya~forest is  $F(\lambda)$, the ratio of the observed flux to the unabsorbed (intrinsic) flux of a background source, as a function of the observed wavelength. In practice, the measured $F(\lambda)$ is affected by various effects, such as uncertainties in the source's continuum, instrumental noise, absorption by metals... We will address these matters at a later stage, and for now, we will consider that $F$ solely reflects the \lya~absorption in the dilute IGM. To study the correlated fluctuations in $F$, we define the $\delta$ field, called density contrast of the \lya~forest, as:
\begin{equation}\label{eq:delta}
    \delta(\boldsymbol{\theta}, \lambda) = \frac{F(\boldsymbol{\theta}, \lambda)}{\overline{F}(\lambda)} -1
\end{equation}

Here $\boldsymbol{\theta}$ represents an angular direction in the sky, i.e.~a line-of-sight. $\lambda$ is an observed wavelength  and $\overline{F}(\lambda)$ is the sky-averaged value of $F$ at $\lambda$, corresponding to the mean IGM absorption at redshift $1 + z = \lambda/\lambda_{\alpha}$. 

While $\delta$ fluctuations exhibit a strong non-Gaussian behavior on small scales, the key statistical information about the density contrast field $\delta$ lies in its two-point correlation function, or equivalently its Fourier transform, the power spectrum. Specifically, the cosmological details pertaining to the underlying matter density field are fundamentally embedded within this statistical function, which is the focal point of interest in this study. We define the correlation function from the ensemble average, as follows:

\begin{equation}\label{eq:xi}
    \xi_{\rm 3D}(z,\theta,\Delta\lambda) \equiv \left < \delta(\boldsymbol{\theta_{i}},\lambda_{i}) \; \delta(\boldsymbol{\theta_{j}},\lambda_{j})\right >
\end{equation}

\noindent Here $\theta$ is the angular separation between two directions $\boldsymbol{\theta_i}$ and $\boldsymbol{\theta_j}$, as the \lya~absorption field is isotropic as a function of sky coordinates. The radial separation is parameterized by the two relations: $\Delta\lambda = \lambda_j - \lambda_i$ and $1+z = (\lambda_i + \lambda_j)~/~2\lambda_{\alpha}$. Alternatively, one can parameterize the radial separation in terms of $\Delta v = c \ln(\lambda_j / \lambda_i)$. Throughout this study, we will consider \lya~forest samples in a narrow redshift slice, so that in Eqn.~\ref{eq:xi} and subsequent equations, $z$ is considered to be identical to its mean $\langle z\rangle$. Subsequently, \p3 is defined as the 3-dimensional Fourier transform of the correlation function:
\begin{equation}\label{eq:P3D}
    P_{\rm 3D}(z,k_{\perp},k_{\parallel}) \equiv \int d^2\theta \; e^{i \boldsymbol{\theta \cdot k_{\perp}}} \int d \Delta \lambda \; e^{i \Delta \lambda k_{\parallel}} \; \xi_{\rm 3D}(z,\theta,\Delta \lambda)
\end{equation}
 Given the statistical isotropy of $\delta$, \p3~does not depend on the direction of the 2D vector $\boldsymbol{k_{\perp}}$, but only on its modulus $k_{\perp}$. As an observational quantity, \p3~has units of $\rm deg^{2}\,$\AA~(respectively $\rm deg^{2}$~km s$^{-1}$), $k_{\perp}$ has units of $\rm deg^{-1}$, and $k_{\parallel}$ has units of \AA$^{-1}$ (respectively s km$^{-1}$, if we use $\Delta v$ instead of $\Delta \lambda$).

\begin{figure}[ht]
\centering
\includegraphics[width=1\textwidth]{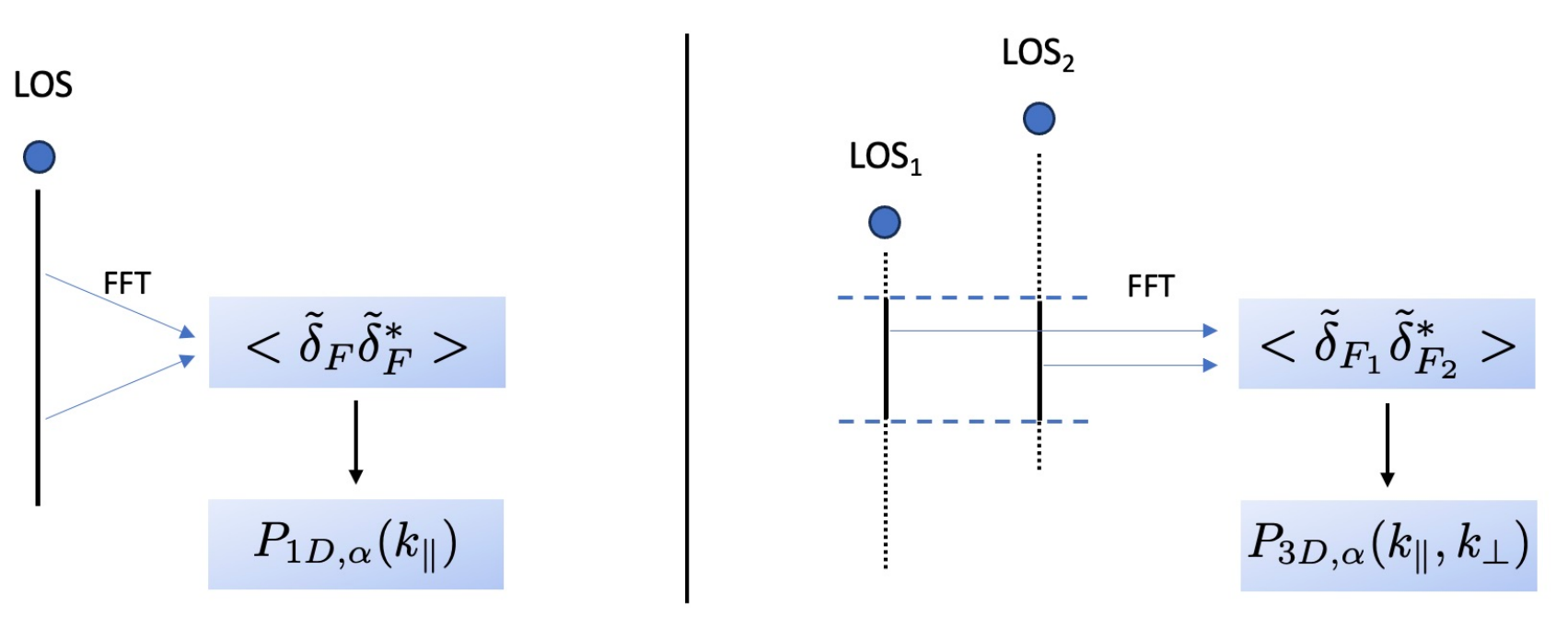}
\caption{Representation of the fundamental geometric principles behind \lya~correlations for the estimation of \p1~and \p3~using a 1D FFT approach. The left panel illustrates the \p1~measurement, derived from a set of FFTs of independent lines-of-sight (LOS), $\Tilde{\delta}_F(k_{\parallel})$. Conversely, the right panel illustrates the \p3~measurement, obtained by correlating 1-dimensional FFTs, namely $\Tilde{\delta}_{F_1}$ and $\Tilde{\delta}_{F_2}$, across different lines-of-sight. This illustration also highlights, through bold lines, the delineation of ``chunks" within \lya~forests, which serve as the basis for correlating distinct FFTs.}
\label{fig:scheme}
\end{figure}

From a technical standpoint, \p3~determines correlations between pixels taken from distinct quasar spectra, as illustrated in Fig.~\ref{fig:scheme}.
On the other hand, \p1~is obtained by correlating pixels belonging to the same quasar spectrum. It is therefore defined as the Fourier transform of the 1-dimensional correlation function $\langle \delta(\boldsymbol{\theta},\lambda_i) \delta(\boldsymbol{\theta},\lambda_j)\rangle$ at fixed $\boldsymbol{\theta}$. It is worth noting that the \p1 can be expressed as a function of the $P_{3D}(z,k_{\perp},k_{\parallel})$ in this manner:
\begin{equation}\label{eq:P1D_P3D}
    P_{\rm 1D}(z,k_{\parallel}) = \int \frac{d^2 k_{\perp}}{(2 \pi )^2} \; P_{\rm 3D}(z,k_{\perp},k_{\parallel})
\end{equation}

\subsection{The cross-spectrum}\label{sec:pcross}

The key element in the approach we propose to measure the \p3, is to first consider a hybrid quantity between real and Fourier space, referred to as the cross-spectrum $P_{\times}(z,\theta, k_{\parallel})$. Similarly to \p3, \px~quantifies longitudinal correlations in Fourier space. However, for transverse correlations, instead of being expressed in terms of the perpendicular wavenumber $k_{\perp}$, the cross-spectrum resembles a correlation function, being defined as a function of the real-space angular separations $\theta$ between the correlated pixels. To our knowledge, \px~was first introduced in the context of the \lya~forest by~\cite{Hui:1998vf}, and also studied in~\cite{Viel:2001hd,Rorai:2013dxa,Font-Ribera:2017txs}. Here, we use the same notation \px~as in~\cite{Hui:1998vf,Font-Ribera:2017txs}. However, let us emphasize that \px~is unrelated to the full Fourier space power spectrum computed from the cross-correlations between two distinct fields. \px~can be expressed both as a function of $\xi_{3D}$ or \p3 in the following way:
\begin{eqnarray}\label{eq:Pcross_1}
    P_{\times}(z,\theta, k_{\parallel}) & \equiv & \int \; d\Delta \lambda \; e^{i\Delta \lambda k_{\parallel}} \; \xi_{3D}(z,\theta,\Delta \lambda) \\
     & = & \int \frac{d^2 k_{\perp}}{(2 \pi)^2} \; e^{i\boldsymbol{\theta}\cdot\mathbf{k_{\perp}}} \; P_{3D}(z,k_{\perp},k_{\parallel})
\end{eqnarray}

\noindent The symmetries of the problem, and the assumption that $z$ is fixed, ensure that \px~is a real quantity. Then, \p1 as defined in Eqn.~\ref{eq:P1D_P3D}, can be seen as a special case of \px, for null angular separation:
\begin{equation}\label{eq:P1D_Pcross}
    P_{1D}(z,k_{\parallel}) = P_{\times}(z,\theta=0,k_{\parallel})
\end{equation}

\px~is a correlator well-adapted to the geometry of \lya~forest observations. The measured \lya~absorption ``pixels'' are aligned along individual lines-of-sight, usually on a common dense grid of wavelengths: this makes it easy to use the Fourier representation along the parallel direction. On the other hand, the lines-of-sight are sparsely and inhomogeneously distributed in sky coordinates, so that a real-space representation of their correlations is more straightforward.

Defining $\Tilde{\delta}(\boldsymbol\theta_i,k_{\parallel})$ as the Fourier transform of each individual line-of-sight $\delta({\boldsymbol\theta_i},\lambda)$, the definition given by Eqn.~\ref{eq:Pcross_1} implies that:

\begin{equation}\label{eq:Pcross_2}
    \left < \Tilde{\delta}(\boldsymbol\theta_i,k_{\parallel}) \;  \Tilde{\delta}^*(\boldsymbol\theta_j,k_{\parallel}')\right > = 2\pi \; \delta^{D}(k_{\parallel}-k_{\parallel}') \; P_{\times}(\theta,k_{\parallel})
\end{equation}

\noindent where $\delta^D$ is the Dirac delta function and $\theta$ again represents the angular separation between the $\boldsymbol\theta_i$ and $\boldsymbol\theta_j$ directions.

\subsection{A fast and simple estimator for \texorpdfstring{$P_{\times}$}{P\_x}}\label{sec:estimator}

Considering Eqn.~\ref{eq:Pcross_2}, it becomes evident that we can expand the 1-dimensional FFT approach, which has already been employed to compute \p1 as seen in~\cite{Chabanier:2018rga,DESI:2023xwh}, to also compute \px.

In practice, the starting quantity for our \px~measurement at a given redshift $z$, is a set of density contrasts $\delta(\boldsymbol{\theta_i}, \lambda)$, extracted from the quasars' \lya~forests in a data sample. Importantly, we assume here that all the $\delta_i$ are tabulated on a common regular wavelength grid:
\begin{itemize}
    \item A common wavelength binning, which is the case of spectra published e.g.~by SDSS and DESI. If this is not achieved, one would additionally need to handle phase shifts in the $\Tilde{\delta}_i\,\Tilde{\delta}^*_j$ products.
    \item A common wavelength range, centered at $(1+z)\,\lambda_{\alpha}$: indeed, for each pair of lines-of-sight $(i,j)$, the wavelength ranges of $i$ and $j$ must be the same to compute the products of Fourier transforms.
\end{itemize}
\noindent A consequence of these requirements is that only a fraction of each quasar's \lya~forest is used, depending on each quasar's redshift, as illustrated in Fig.~\ref{fig:scheme}. This, results in a loss of statistics. However, such a choice was not needed in the case of \p1, e.g. in~\cite{Chabanier:2018rga} the density contrasts $\delta_i$ were computed on ``chunks'' of \lya~forests defined in each quasar's rest-frame, so that all the selected \lya~forest samples contributed to the measurement.

Employing a Fast Fourier Transform method, we initially perform a 1-dimensional Fourier transform on all density fields: $\delta(\boldsymbol{\theta_i},\lambda) = \delta_i(\lambda) \xrightarrow{FFT} \Tilde{\delta}(\boldsymbol{\theta_i},k_{\parallel}) = \Tilde{\delta}_i(k_{\parallel})$. Following this, we calculate the angular separations $\theta_{ij}$ for all possible pairs of lines-of-sight $(i,j)$, and we bin these angular separations, depending on the angular separation values for which we intend to calculate \px: an accurate determination of \px~is possible only if there is a sufficient number of line-of-sight pairs within each respective bin. For each selected bin, we define the average angular separation $\theta$ as the mean value of $\theta_{ij}$ corresponding to the pairs that contribute to that specific bin.
For a given angular separation bin, considering the sample of line-of-sight pairs $(i,j)$ falling in this specific bin, we compute all products of 1-dimensional FFTs, $\Tilde{\delta}_i\,\Tilde{\delta}^*_j$. Subsequently, by averaging over all pairs, one can obtain an estimate of \px, as outlined in Eqn.~\ref{eq:Pcross_2}. We define our estimator:

\begin{equation}\label{eq:Pcross_3}
    P_{\times}(\theta,k_{\parallel}) = \left < \Re\left(\Tilde{\delta}_{i}(k_{\parallel}) \; \Tilde{\delta}_{j}^{*}(k_{\parallel})\right) \right >
\end{equation}

Since \px~is real, the average of the imaginary parts of $\Tilde{\delta}_{i}\Tilde{\delta}_{j}^{*}$ products is zero: therefore, to reduce statistical fluctuations we only consider their real parts, $\Re(\Tilde{\delta}_{i}\Tilde{\delta}_{j}^{*})$, in Eqn~\ref{eq:Pcross_3}. This is equivalent to including both $(i,j)$ and $(j,i)$ pairs in the averaging process. In the special case of identical line-of-sight pairs $(i,i)$, i.e.~$\theta$ = 0, this product simplifies to computing the average of $|\delta_i|^2$, which corresponds to the conventional FFT estimator for \p1. 
As for \p1, a set of $P_{\times}(\theta,k_{\parallel})$ can be computed separately for different redshift bins, using only \lya~pixels centered around a mean redshift $\langle z \rangle$.

The parallel wavenumbers $k_{\parallel}$ are dictated by the (common) wavelength grid. In this context, we assume a constant pixel size, which can be either $\Delta \lambda_{\rm pix}$ in \AA, or $\Delta \log \lambda_{\rm pix}$. Defining the number of pixels $N_{\rm pix}$, the minimum non-zero wavenumber is 
$\frac{2\pi}{N_{\rm pix}} \; \Delta\lambda_{\rm pix}$.
Its maximum value equals the Nyquist frequency $\frac{\pi}{\Delta\lambda_{\rm pix}}$. 

We derive the statistical error bars and covariance for \px~within a specific angular separation bin, using the dispersion of individual $\delta_i \delta_j$ measurements. This is identical to the way it is done for \p1~\cite{Chabanier:2018rga}. The underlying assumption is that individual pair measurements $(i,j)$ are independent of each other. This assumption holds in the context of this study, because we are considering small angular separations and using relatively low-density \lya~observations over a substantial portion of the sky: the probability for a line-of-sight $(i)$ to contribute to two pairs $(i,j)$ and $(i,j')$ within the same angular separation bin is quite low. Due to the same reason, we anticipate that the statistical covariance of \px~will be minimal between different angular separation bins. Consequently, it is not taken into account in this study. It is important to note that if a future work expands the measurement of \px~to larger angular separations, a more refined estimation of the covariance would be essential.

In terms of computational requirements, this method demands very limited resources. Some computation time is needed to tabulate pairs of lines-of-sight as a function of angular separations. The subsequent step, which involves computing 1D FFTs for all individual lines-of-sight, requires a computation time identical to the case of \p1. Overall, as an example, our computations indicate that it takes only a few minutes, when executed on a single CPU node of the Perlmutter machine at the NERSC computing center, to compute \px~in a single redshift bin, for the SDSS sample considered in Section~\ref{sec:appli-SDSS}.

\subsection{Noise and resolution}\label{sec:Noise_resolution}
When dealing with real data, given that we are interested in small-scale correlations, the main instrumental effects that need to be corrected for are the spectroscopic resolution and its noise.  Although such corrections depend on the considered data set, they are relatively similar between different instruments. We consider here the case of SDSS spectra, since we later apply our method to SDSS data sets.

The expression of the measured density contrast for a line-of-sight $i$, $\delta_i(\lambda)$, can be written as a function of the ``astrophysical''  contrast $\delta_{i,{\rm Ly\alpha}}(\lambda)$ as follows:
\begin{equation}\label{eq:delta_data_real}
    \delta_i(\lambda) =  \delta_{i,{\rm Ly\alpha}}(\lambda) \ast W_i(\lambda,R) + \delta_{i,n}(\lambda)
\end{equation}

\noindent In this expression, $\delta_{i,{\rm Ly\alpha}}$ is convolved by the kernel function $W_i(\lambda,R)$ that accounts for the spectral response of the spectrograph, as well as the sample's wavelength pixelization.
Here, as in the \p1~measurement of \cite{Chabanier:2018rga} using SDSS data, we approximate $W$ by using a mean resolution parameter $R$ for each measured sample.
$\delta_{i,n}(\lambda)$ represents noise fluctuations: those are usually well represented by a Gaussian white noise model, whose amplitude depends on the considered spectrum. In Fourier space, Eqn.~\ref{eq:delta_data_real} becomes:
\begin{equation}\label{eq:delta_data_fourier}
    \Tilde{\delta}_i(k_{\parallel}) = \Tilde{\delta}_{i,{\rm Ly\alpha}}(k_{\parallel}) \times \widetilde{W}_i(k_{\parallel},R)  + \Tilde{\delta}_{i,n}(k_{\parallel})
\end{equation}

\noindent We define $P_{\times,{\rm Ly\alpha}}$ and \px~as the cross-spectra of $\delta_{i,{\rm Ly\alpha}}(\lambda)$ and $\delta_i(\lambda)$
respectively. We also define the noise (cross) power spectrum in the following way:

\begin{equation} \label{eq:P_noise_general}
    P_n(\theta, k_{\parallel}) = \langle \Re(\Tilde{\delta}_{i,n}\Tilde{\delta}^*_{j,n}) \rangle\
\end{equation}

\noindent In this expression, similarly to previous expressions, $(i,j)$ pairs are separated by $\theta$. It is then straightforward to derive $P_{\times,{\rm Ly\alpha}}$ from the observed \px, by employing an equation that is an extension of the one used for \p1, e.g. in~\cite{Chabanier:2018rga}:

\begin{equation}\label{eq:P_lya}
    P_{\times,\rm Ly\alpha}(\theta, k_{\parallel}) = \frac{P_{\times}(\theta, k_{\parallel}) - P_{n}(\theta, k_{\parallel})} {\left <\widetilde{W}_i(k_{\parallel}) \, \widetilde{W}_j(k_{\parallel}) \right > }
\end{equation}

Since the noise is independent of the \lya~signal, all $\langle \Tilde{\delta}_{i,n} \Tilde{\delta}_{j,{\rm Lya}}^{*} \rangle$ vanish. 
If we neglect the impact of correlated noise between different measured spectra, we have $\langle \Tilde{\delta}_{i,n} \Tilde{\delta}^{*}_{j,n} \rangle = 0$ for any pair $(i,j)$ of lines-of-sight with $i \neq j$. We will come back to this assumption in the specific case of SDSS spectra in Section~\ref{sec:eBOSS_measurement}. In that case, the only non-zero noise correlation terms are for $i=j$, therefore we have:

\begin{equation}\label{eq:P_noise}
P_n(\theta, k_{\parallel}) = 
\begin{cases}
 \langle |\Tilde{\delta}_{i,n}|^2 \rangle & \text{if $\theta = 0$} \\
 ~~~~0 & \text{if $\theta > 0$} \\
\end{cases}
\end{equation}

\noindent Hence, contrarily to the case of \p1, there is no noise correction to apply for $\theta>0$.

\subsection{From \texorpdfstring{$P_{\times}$}{P\_x} to \texorpdfstring{$P_{3D}$}{P\_{3D}}}\label{sec:p3d}

The measurement of \px~itself is of interest, and as we will emphasize in the conclusion, it could certainly be directly used for cosmological inferences within a full modeling approach. However, the computation of $P_{\times}(z,\theta,k_{\parallel})$ also allows one to infer \p3. Using Eqn.~\ref{eq:P3D}, we have:
\begin{equation}\label{eq:P3D_Pcross_1}
    P_{\rm 3D}(z,k_{\perp},k_{\parallel}) = \int d^2\theta \; e^{i \boldsymbol{\theta \cdot k_{\perp}}} \; P_{\times}(z,\theta, k_{\parallel}) 
\end{equation}

\noindent Moving to cylindrical coordinates, the angular symmetry of \px~permits this 2D integral to be simplified to a 1D integral:

\begin{equation}\label{eq:P3D_Pcross_3}
        P_{\rm 3D}(z,k_{\perp},k_{\parallel}) = 2\pi \int_0^{\infty} d\theta \, J_{0}(k_{\perp}\theta) \,\theta \, P_{\times}(z,\theta,k_{\parallel})
\end{equation}

\noindent Here $J_{0}(x)$ is a first kind Bessel function of order zero, which oscillates while slowly decreasing as a function of $x =  k_{\perp}\theta$.

Following Eqn.~\ref{eq:P3D_Pcross_3}, one straightforward approach for inferring \p3~involves performing numerical integration of \px, which is the technique used in this work. Numerically, since \px~is expected to be a smooth function of $\theta$ for the scales of interest, we perform an interpolation of \px~over a finely-spaced grid of $\theta$ values, using a smoothing spline fit. This ensures that we accurately capture the oscillations of the Bessel function $J_{0}(x)$. Using the interpolated cross-spectrum, we perform the numerical integration using Eqn.~\ref{eq:P3D_Pcross_3}.
This integration can be done only for certain values of $k_{\perp}$ and $k_{\parallel}$. First of all, $\theta_{\rm min}$, the smallest non-zero angular separation for which we measure \px, dictates the largest perpendicular mode $k_{\perp,{\rm max}} \propto 1/\theta_{\rm min}$. In addition to that, the numerical integration makes sense only if typical variations of \px~as a function of $\theta$ take place on scales larger than $\theta_{\rm min}$. In practice, \px~is a rapidly decreasing function of $\theta$ since it is a correlation function as a function of transverse coordinates. Therefore, a simple and somewhat arbitrary criterion we choose is to require that $P_{\times}(\theta_{\rm min}) / P_{\times}(\theta=0) > 0.3$. Given that \px~decreases faster with $\theta$ for larger values of $k_{\parallel}$, this results in a maximal value $k_{\parallel,{\rm max}}$. Using a simple toy model for \px, we estimate that the systematic error on \p3 due to the finite value of $\theta_{\rm min}$ is of the order of 20~\%, with this criterion and our chosen integration method.

Within this method, the \p3~statistical uncertainties are propagated following a simple Monte-Carlo approach. Initially, random realizations of \px~are drawn according to the measured \px~covariance matrix. Subsequently, the statistical errors on the measured \p3~are given by the dispersion of the random \p3~values obtained after numerical integration of the random \px~realizations, according to Eqn.~\ref{eq:P3D_Pcross_3} as well.

It is evident that there exist more advanced approaches for estimating \p3, such as implemented in~\cite{Font-Ribera:2017txs}. This will be discussed later in the conclusion. Here we use our simple numerical integration due to the fact that the choice of an alternative \p3~estimator depends on the strategy used to physically interpret the measurement - a step that we leave for a future work. We emphasize that regardless of the chosen method, the derivation of \p3~from \px~demands negligible computational resources in comparison to the \px~computation itself. Additionally, we highlight that the numerical integration approach is completely model-independent, which is not the case for other approaches such as the quadratic estimator used in~\cite{Font-Ribera:2017txs}, in which \p3~is parameterized using an expansion around a reference model, or even more using explicit fitting functions such as those of~\cite{Arinyo-i-Prats:2015vqa}.

\paragraph{Units}\label{sec:Units_conversions}
\begin{itemize}
\item When modelling the \lya~forest, for example in the case of simulated data derived from hydrodynamical simulations detailed in Section~\ref{sec:NyxBoxValidation}, with a known background cosmology, \px~and \p3~are computed in \mpcph~and [\mpcph]$^3$ respectively. Angular separations $\theta$ are also expressed in \mpcph. This allows one to compare measured power spectra to the ``truth'' power spectra of the simulations, whose coordinates are given in comoving Mpc.

\item When using real data, it is more appropriate to provide a cosmology-independent measurement, that can be later interpreted within any cosmological model. As already mentioned in Section~\ref{sec:method}, depending on whether the pixel size along lines-of-sight is expressed in \AA~or in km s$^{-1}$ (in the case of a log scale wavelength grid), \px~is expressed in \AA~or km s$^{-1}$ units, and \p3 is expressed in [deg$^2$~\AA] or in [deg$^{2}$~km~s$^{-1}$].
\end{itemize}

\noindent For reference, in the case of a flat cosmological model with comoving distance $D_c(z)$ in Mpc, and Hubble rate $H(z)$ in km~s$^{-1}$~Mpc$^{-1}$, and for data sets at mean redshift $z$, with $c$ in km~s$^{-1}$ and $\lambda_{\alpha}$ in \AA, the conversion formulae are:
\begin{eqnarray}\label{eq:P_conversions}
   P_{\times} [h^{-1}\mathrm{Mpc}] & = & \frac{h c}{\lambda_{\alpha}~H(z)}~P_{\times} [\text{\AA}] \\
   P_{\times} [\mathrm{km~s^{-1}}] & = & \frac{c}{\lambda_{\alpha}~(1+z)}~P_{\times} [\text{\AA}] \\
   P_{\rm 3D} [h^{-1}\mathrm{Mpc}]^3 & = & \left(\frac{\pi}{180}\right)^2\frac{h^3\,c}{\lambda_{\alpha}} \frac{D_c(z)^2}{H(z)}~P_{\rm 3D}[\mathrm{deg^2}~\text{\AA}]
\end{eqnarray}

\section{Validation with hydrodynamical simulations}\label{sec:NyxBoxValidation}

In this section, we apply the previously outlined method to simulated \lya~absorption fields. We make use of the output of a cosmological simulation, run with the \nyx~software. This provides tests of our measurement strategy with a realistic \lya~forest model, and more importantly, a well-defined input \p3.

\subsection{Description of the simulations}\label{sec:description}
Simulated \lya~forest samples have been generated over large volumes such as those covered by large spectroscopic surveys, e.g. in~\cite{Bautista:2014gqa,Farr:2019xij}. However, such mock samples are optimized to mimic large-scale correlations, and do not necessarily reproduce a well-defined small-scale \p3 of the \lya~absorption field. On the other hand, using a simple Gaussian random field with a known \p3 may be unrealistic, since at small scales, the \lya~absorption field is highly non-Gaussian. We therefore resort to the use of \lya~absorption fields as computed by cosmological hydrodynamical simulations. Currently, these simulations cannot span over a cosmological volume equivalent to the one probed by BOSS, but we can compensate for this limitation by drawing lines-of-sight inside the simulation box with an appropriate density, ensuring that the statistics of relatively small-separation pairs reasonably matches that of real data.

\nyx~\cite{Almgren:2013sz,Sexton:2021xea} is a cosmological simulation code, solving the evolution of the baryonic gas coupled to dark matter in the expanding Universe. While dark matter is treated with an N-body approach, the gas hydrodynamics are solved on a mesh. \nyx~is particularly well adapted to model the \lya~forest, as described in details in~\cite{Lukic2014}, and applied in e.g.~\cite{Walther:2020hxc,Chabanier:2022dgr}. The Eulerian, fixed-mesh algorithm turns out to be efficient to model the small-scale fluctuations of baryon properties in the low-density regions of the cosmic web, where most of the \lya~absorption signal is produced. The \lya~forest modelling is done by following electrons, hydrogen and helium species (neutral and ionized), making use of relevant atomic, heating and cooling processes. Concerning the post-processing of a \nyx~simulation, this is done using the \texttt{Gimlet}~software~\cite{Briesen2016}. \texttt{Gimlet} is a C++, MPI-parallel toolkit well adapted to \nyx~outputs~\cite{Lukic2014}.

\begin{figure}[ht!]
    \centering
    \includegraphics[width=1\columnwidth]{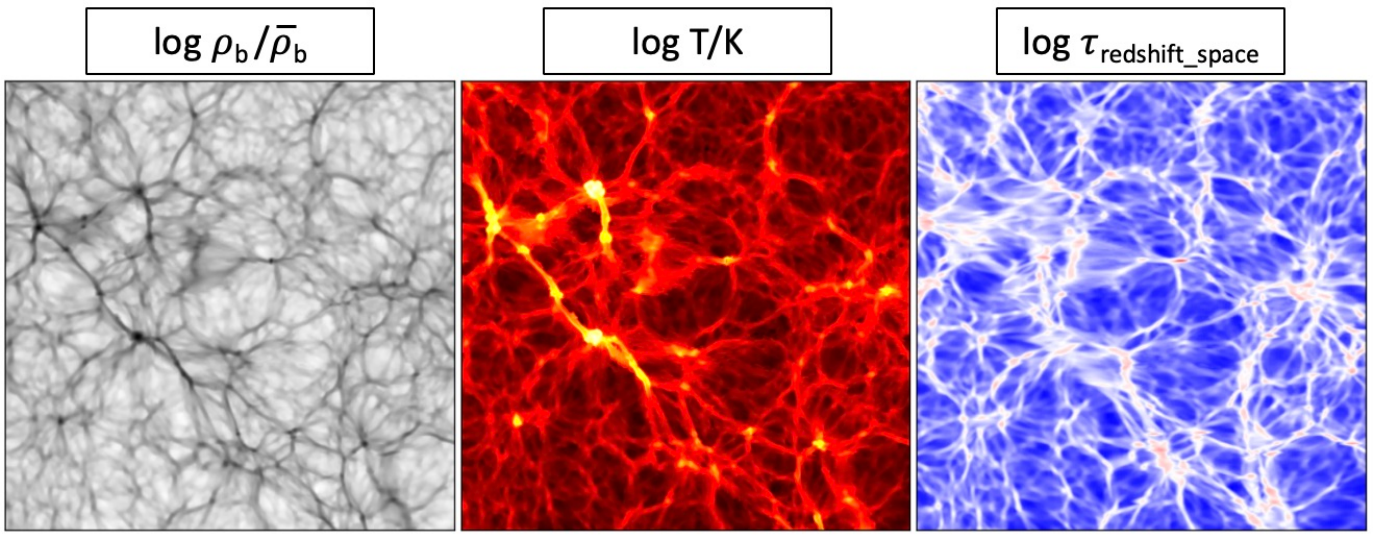}
    \caption{Slice of the baryon density, temperature, and optical depth in redshift space from the employed \nyx~simulation, with 1536$^3$ cells and box size of 150 \mpcph. The baryon density and temperature fields are statistically isotropic. For the optical depth field, the horizontal axis corresponds to the line-of-sight: the effects of IGM temperature, peculiar velocities and redshift space distortions result in anisotropies of its statistical properties.}
\label{fig:NyxSlices}
\end{figure}

Here we make use of an output (snapshot) from a \nyx~simulation that is described in more details in~\cite{Chabanier:inprep}. The simulation was run starting from $z=200$, with Zel'dovich initial conditions, and cosmological parameters from~\cite{Planck:2015fie}: $\Omega_b = 0.0487$, $\Omega_m = 0.31$, $H_0 = 67.5$, $n_s = 0.96$ and $\sigma_8 = 0.83$. Importantly for the \lya~forest, a spatially uniform UV+X-ray radiation field is assumed to dictate the ionization state of Hydrogen and Helium, following the middle reionization scenario of~\cite{Onorbe:2016hjn}. The simulated cosmological volume is a large cubic box, $150$~\mpcph~wide, with $1536^3$ gas cells, resulting in a resolution of $98$~\kpcph. This volume is large enough to draw lines-of-sight with an appropriate number density. Additionally, the spatial resolution is good enough to physically capture some of the relevant small-scale features in the \lya~forest. Note that it is not enough to fully resolve the Jeans scale $\lambda_J \sim 80$~\kpcph, but this is not important for this study: the minimum non-zero angular separation bin $\theta_{\rm min}$ we use in mocks with a ``realistic'' line-of-sight density, and with SDSS data, corresponds to transverse distances $\sim 0.6$~\mpcph, much larger than $\lambda_J$. For practical reasons we use the simulation output at redshift $z=2.0$.

The IGM's \lya~optical depth $\tau_{\alpha}$ is numerically calculated in redshift space from simulation outputs using \texttt{Gimlet}, which takes into account a Gaussian thermal line broadening. A slice of $\tau_{\alpha}$, computed in redshift space with respect to the line-of-sight, is shown on the right-hand side of Fig.~\ref{fig:NyxSlices}. We also show the corresponding baryon density and temperature fields, that were used to compute $\tau_{\alpha}$. These density and temperature fields are statistically isotropic. Nonetheless, the effects of thermal broadening, peculiar velocities and redshift space distortions have a clear impact on $\tau_{\alpha}$, resulting in statistical anisotropies in this field, with a preferred axis given by the line-of-sight.

\subsection{Mock generation}\label{sec:mock_generation}

In order to validate our measurement approach, we generate mock line-of-sight samples from the simulations described in Section~\ref{sec:description} and perform measurements of \px~and \p3.
Starting with the computed $\tau_{\alpha}$ grid, we derive the transmitted flux fraction grid, $F(\lambda) = e^{-\tau_{\alpha}}$. Additionally, we compute the corresponding grid of density contrasts $\delta$ of the \lya~forest based on Eqn.~\ref{eq:delta}. Following this, a 3D \lya~absorption power spectrum \p3~is computed using \texttt{Gimlet}, which relies numerically on a 3D FFT, well-adapted to the regularly-gridded $\delta$ field. We later refer to this \p3~as the ``truth'', although it is essential to acknowledge that this power spectrum is also altered by numerical effects, stemming from the FFT process, as well as the effect of cosmic variance due to the finite box size.

Line-of-sight mock samples are finally generated from the $\delta$ grid. To create them, we initially select random floating-point values for the $x$ and $y$ coordinates of each line-of-sight, and subsequently determine the corresponding $\delta$ values along the z-axis. To be more precise, the $\delta$ of each line-of-sight of random $x$ and $y$ coordinates is obtained through a linear interpolation on the $\delta$ grid. This method ensures that the lines-of-sight are randomly distributed, even when dealing with small angular separations. By construction, the pixels of individual lines-of-sight are all on the same grid, with a $98$~\kpcph~pixel size.

\subsection{Results with a reference, high-density and noiseless mock}\label{sec:validation_Nyx}

\begin{figure}[ht]
\centering
\includegraphics[width=0.79\textwidth]{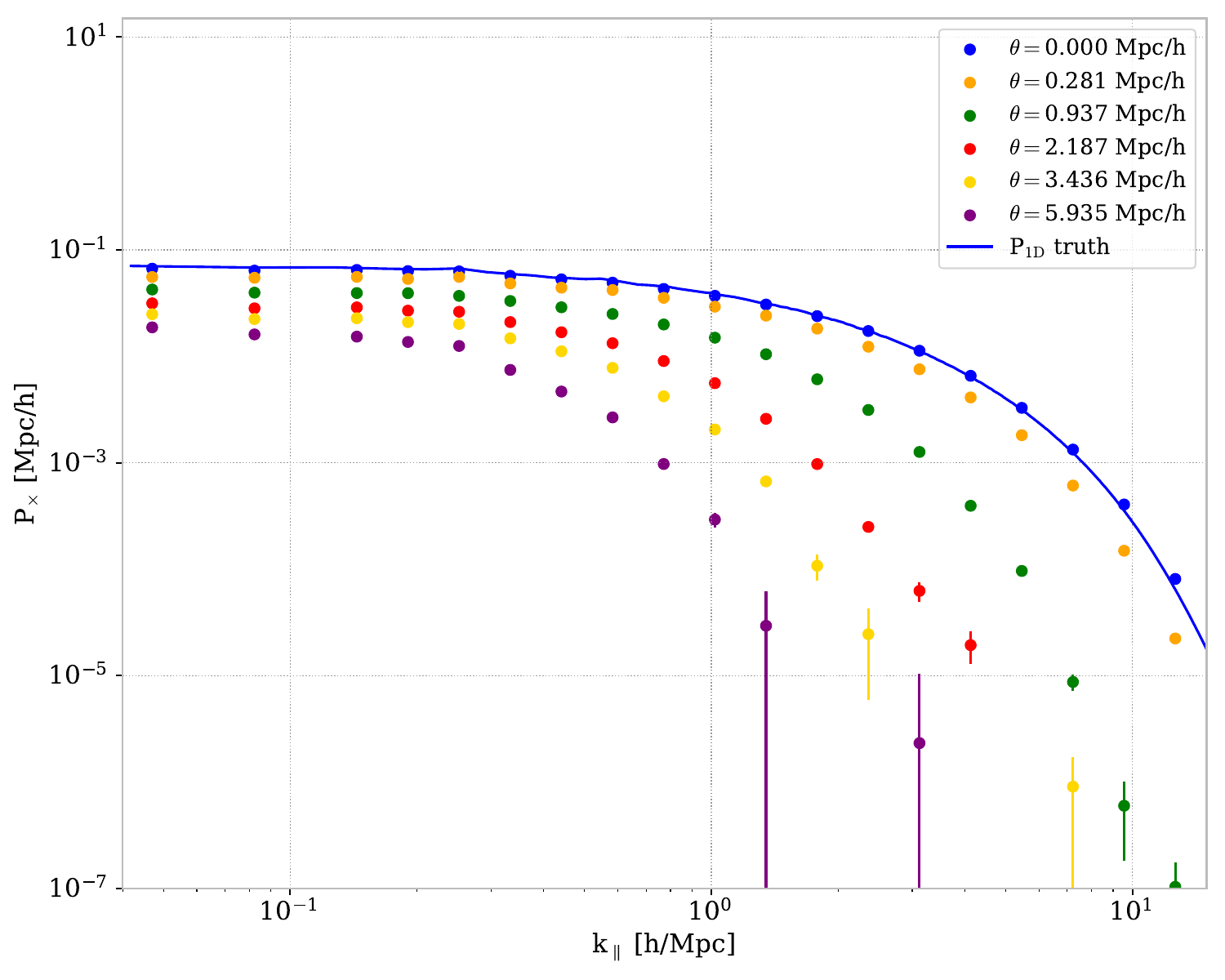}
\caption{Estimated \px$(\theta,k_{\parallel})$ from our reference mock of $10^4$ noiseless lines-of-sight drawn from the 150~\mpcph~\nyx~box at z = 2.0. $P_{\times}(k_{\parallel}, \theta=0)$ being the \p1, is compared to the ``truth" \p1~provided by \texttt{Gimlet} and represented by a continuous blue line. We rebinned \px~values on a logarithmic grid in $k_{\parallel}$. Error bars indicate statistical uncertainties.}
\label{fig:Pcross_10000}
\end{figure}

As a first validation of our pipeline, we build a reference mock sample with $10^4$ lines-of-sight, which does not include noise nor resolution effects. To compute \px~on this mock, we follow the procedure outlined in Section~\ref{sec:method}. First, after having computed the transverse comoving separations, and corresponding angular separations of all possible line-of-sight pairs in our sample, we choose angular separation bins centered at $\langle \theta \rangle =$ 0$^{\circ}$ (case of \p1), 0.0015$^{\circ}$, 0.0045$^{\circ}$, 0.008$^{\circ}$, followed by a regular binning of width 0.01$^{\circ}$ ranging from 0.015$^{\circ}$ to 0.35$^{\circ}$, and finally the two bins 0.45$^{\circ}$ and 0.55$^{\circ}$. The maximum considered value is 0.55$^{\circ}$, equivalently 35~\mpcph, much smaller than the size of the simulation box. Then, for each angular separation bin, we proceed with the measurement of $P_{\times}(\theta,k_{\parallel})$ which is represented in Fig.~\ref{fig:Pcross_10000} for selected angular separation bins.

\begin{figure}[ht!]
\centering
\includegraphics[width=0.79\textwidth]{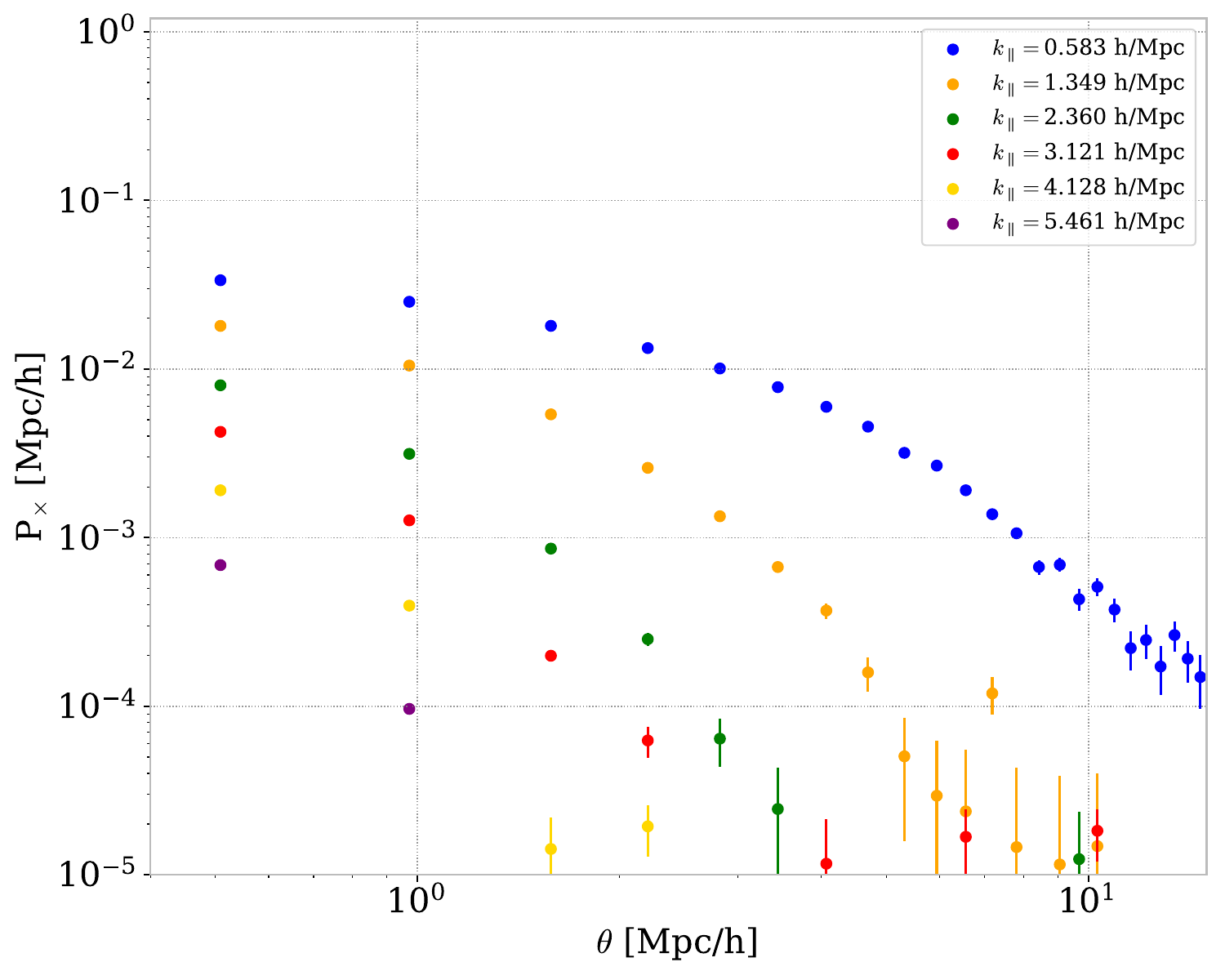}
\caption{Same as Fig.~\ref{fig:Pcross_10000}, but representing \px~as a function of angular separation $\theta$, i.e.~highlighting the fact that \px~is a correlation function as a function of transverse separation. Notably, the transverse correlations in the $\delta$ field decrease faster as a function of $\theta$ when considering short-wavelength longitudinal modes (large $k_{\parallel}$) with respect to long-wavelength modes.}
\label{fig:P_cross_r_10000LOS}
\end{figure}

Initially, we checked that the measured $P_{\times}(k_{\parallel}\,|\,\theta=0)$ is in agreement with the ``truth" \p1 computed by \texttt{Gimlet}. For $\theta > 0$, the shape of $P_{\times}(k_{\parallel})$ is similar to \p1; however when $\theta$ increases, the overall power decreases. Furthermore, there is an attenuation of $P_{\times}$ as a function of $k_{\parallel}$ and this effect intensifies at higher $\theta$ values. In order to highlight this feature, Fig.~\ref{fig:P_cross_r_10000LOS} shows the same measurement, but as a correlation, i.e.~as a function of $\theta$: as expected and already highlighted in e.g.~\cite{Viel:2001hd}, the transverse correlations in the $\delta$ field decrease faster when considering short-wavelength longitudinal modes (large $k_{\parallel}$) with respect to long-wavelength modes.

\begin{figure}[ht]
\centering
\includegraphics[width=0.79\textwidth]{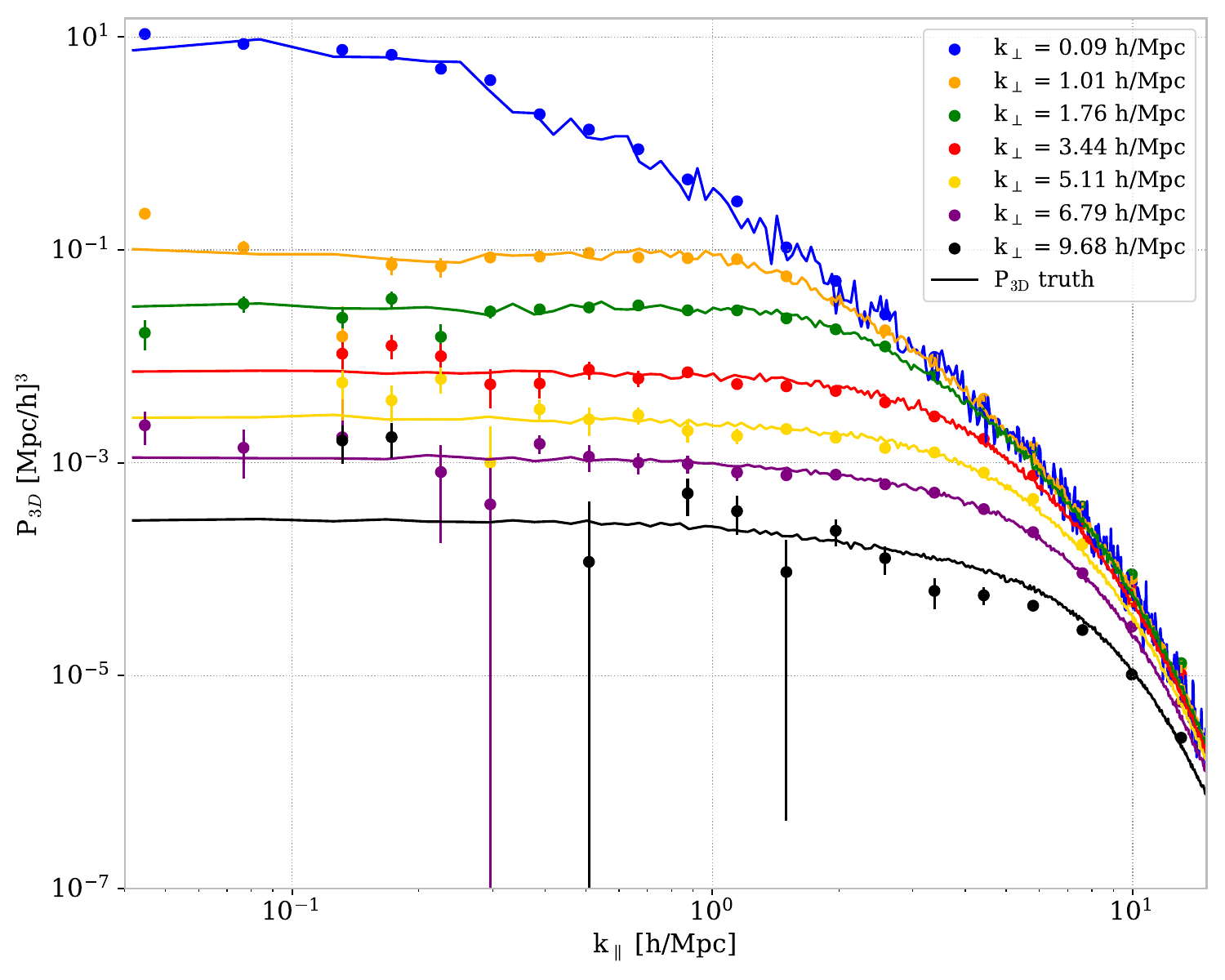}
\caption{Estimated \p3$(k_{\perp},k_{\parallel})$ from our reference mock of $10^4$ noiseless lines-of-sight within the 150 \mpcph~\nyx~box at $z = 2$. Continuous lines show the ``truth" \p3~provided by \texttt{Gimlet}, estimated by full 3D FFT of the box field. As in Fig.~\ref{fig:Pcross_10000}, we rebinned our \p3 values on a logarithmic grid in $k_{\parallel}$, and error bars represent statistical uncertainties.}
\label{fig:P3D_10000}
\end{figure}

We then derive $P_{3D}(k_{\perp},k_{\parallel})$, for values of $k_{\perp}$ identical to those tabulated in the ``truth'' \p3~computed by \texttt{Gimlet}. Fig.~\ref{fig:P3D_10000} presents the computed \p3 for selected $k_{\perp}$ values ranging from~$\sim 0.1$ to~$\sim 10\,h\,{\rm Mpc}^{-1}$, as well as the corresponding ``truth''. The agreement between both estimations is very good: this validates our method to estimate \p3 for a large range of $(k_{\perp}, k_{\parallel})$. In this ideal setup, the number of line-of-sight pairs used to compute \px~is huge, leading to remarkably small statistical error bars. Nonetheless, it is noteworthy that our \p3 estimator exhibits reduced precision for $k_{\perp} \gtrsim 10 \,h\,{\rm Mpc}^{-1}$. This maximal value of $k_{\perp}$ is dictated by the smallest angular separation bin $\theta$.

\begin{figure}[ht]
\centering
\includegraphics[width=0.79\textwidth]{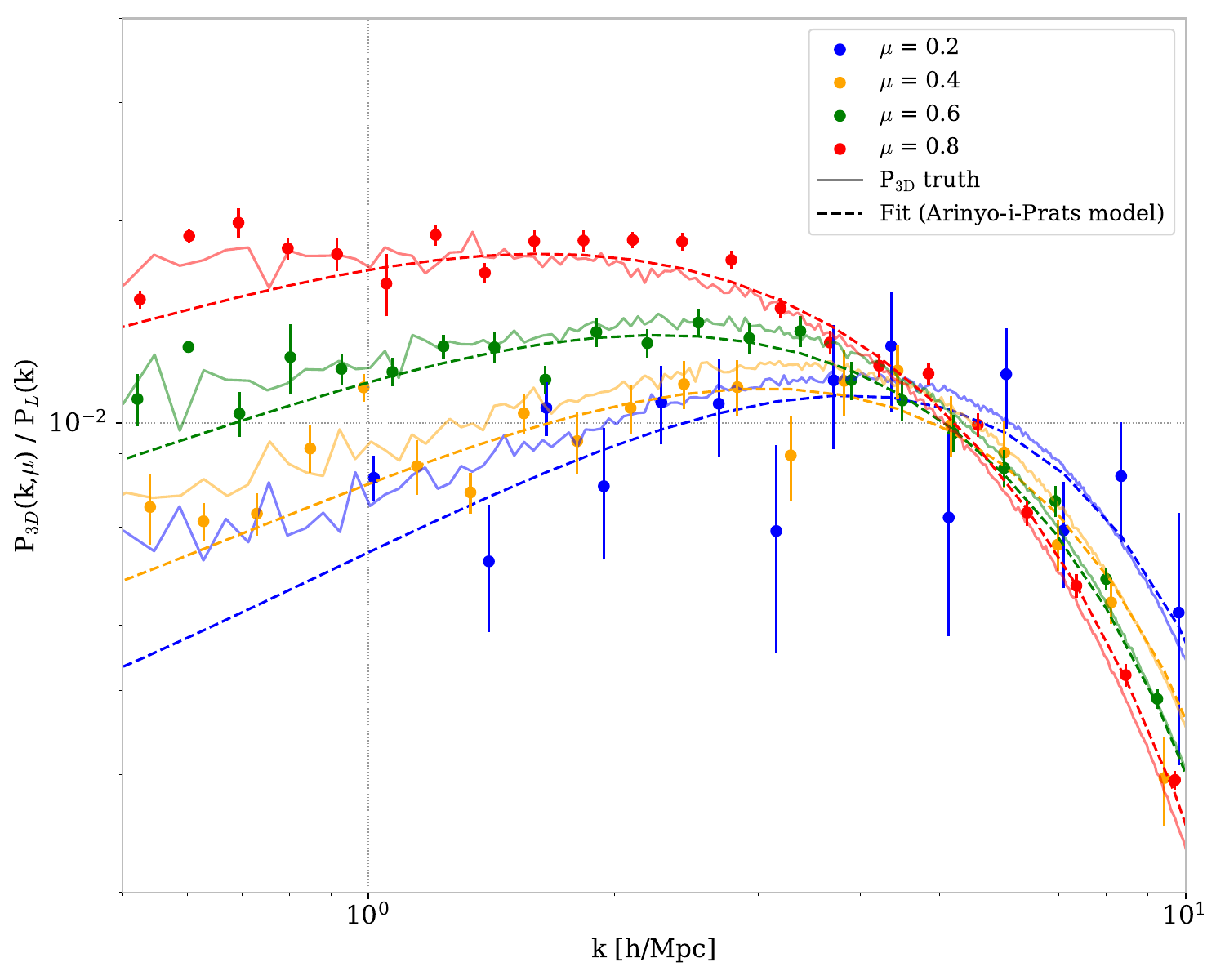}
\caption{Ratio of \p3$(k,\mu)$ to the linear matter power spectrum $P_{\rm L}(k)$, computed for 4 values of $\mu = k_{\parallel}/k$ and as a function of $k = \sqrt{k_{\parallel}^2 + k_{\perp}^2}$. Points with error bars represent the measured \p3~from our reference mock, which is also shown in Fig.~\ref{fig:P3D_10000}, but in a different coordinate system. Continuous thin lines represent the simulation ``truth'' similarly to Fig.~\ref{fig:P3D_10000}. Dashed lines are the result of an 8-parameter fit of the measurement points to Eqn.~\ref{eq:P3D_analytic}, as described in the text.}
\label{fig:P_Pl_10000}
\end{figure}

In order to physically interpret the measured \p3, we perform a change of  coordinates from Cartesian $(k_{\perp},k_{\parallel})$ to polar $(k = \sqrt{k_{\perp}^2+k_{\parallel}^2}, \mu=k_{\parallel}/k)$. We then compute the ratio of \p3 to the (known) linear matter power spectrum $P_{\rm L}(k,z=2)$ estimated by the \texttt{CLASS}~Boltzmann solver~\cite{Blas:2011rf}. This recast of the measurement is shown in Fig.~\ref{fig:P_Pl_10000}. In order to highlight its interpretation, we fit these points with the analytic function given by Eqn.~3.6 of~\cite{Arinyo-i-Prats:2015vqa}:

\begin{equation}\label{eq:P3D_analytic}
    \frac{P_{\rm 3D}(k,\mu)}{P_{\rm L}(k)} = b^2 \, (1+\beta \mu^2)^2 \, \exp\left( (q_1 \Delta^2(k) + q_2 \Delta^4(k)) \left[ 1-\left(\frac{k}{k_v}\right)^{a_v} \mu^{b_v}\right] - \left(\frac{k}{k_p}\right)^2 \right)
\end{equation}

\noindent In this expression $\Delta^2(k) = k^3 P_{\rm L}(k)/2\pi^2$. At low $k$, \p3 is in a near-linear regime: its amplitude is dictated by the bias $b$ and its $\mu$ dependence is driven by the linear RSD term $(1+\beta\mu^2)^2$. Then, the isotropic boost of the power spectrum due to non-linear gravitational growth is parameterized by the $q_1$ and $q_2$ terms. Line-of-sight broadening produced by non-linear peculiar velocities and the IGM temperature generates a strongly $\mu$-dependent cutoff in \p3 ($k_v$, $a_v$, $b_v$ terms). Finally, at large $k$, the impact of Jeans smoothing is isotropic, modelled by the $\exp-(k/k_p)^2$ term. When leaving all parameters free, we derive from our fit \footnote{For this fit, the statistical uncertainties and correlations between parameters are large, so their values should be considered only as indicative.} $(b,~\beta,~q_1,~q_2,~k_v,~a_v,~b_v,~k_p) = (0.043,~1.7,~2.3,~-0.5,~5.3,~0.42,~1.16,~9.4)$. Importantly, the recovered \p3~is in good agreement with the simulation's truth, also displayed in Fig.~\ref{fig:P_Pl_10000}. We notice that for relatively low $k\sim1.5-4$~\mpcph, our measurement systematically overestimates \p3 for large $\mu$, and underestimates \p3 for low $\mu$. This generates a bias on several parameters when fitting Eqn.~\ref{eq:P3D_analytic}. Still, we compared the fit results from our measurement to those from \texttt{gimlet}'s ``truth'', and found that all parameters are within $1.1\,\sigma$. More importantly, we also observe that even in the case of this reference mock sample, the statistical error bars are large when it comes to measure \p3 for small values of $\mu \lesssim 0.3$. This is again related to the limited statistics of very small-separation pairs of lines-of-sight.
In spite of these limitations, it is clear that our \p3 measurement from this reference mock allows us to disentangle the different effects included in the model of Eqn.~\ref{eq:P3D_analytic} - thereby improving our interpretation of \lya~data, both in terms of precision and robustness.

\subsection{Results with realistic mocks}\label{sec:validation_realistic}

In this section, we assess the performance of our measurement approach using realistic mock data sets that resemble existing or forthcoming large spectroscopic quasar samples.~We specifically take into consideration two crucial factors that inevitably influence the accuracy of our measurements:~instrumental noise and lines-of-sight statistics.~A study involving more realistic mocks which encompass elements such as astrophysical contaminants and more detailed instrumental effects, falls beyond the scope of this article.

We first construct a mock sample with the same line-of-sight statistics as in the reference mock discussed in Section~\ref{sec:validation_Nyx}. However, we inject noise into the measured \lya~density contrast $\delta$. This noise component is an uncorrelated, Gaussian white noise with standard deviation $\sigma_{\delta}= 0.5 / \sqrt{\Delta \lambda}$, where $\Delta \lambda$ is the forest pixel size in \AA. This particular value aligns with the typical characteristics of SDSS \lya~samples. To derive \px~from this mock sample, we subtract the noise spectrum $P_n$ when $\theta=0$, as explained in Section~\ref{sec:Noise_resolution}. 

Next, we construct a mock realization with noise-free $\delta$ samples as in the reference mock detailed in Section~\ref{sec:validation_Nyx}. However, in this case, we opt for a reduced number of lines-of-sight, specifically selecting 500 lines-of-sight. We choose this number such that the pairs count at small angular separations resembles those from the SDSS sample given in Table~\ref{tab:Stat_eBOSS} of Section~\ref{sec:appli-SDSS}. For instance, this mock contains 70 line-of-sight pairs contributing to the angular separation bin $\theta=0.03^{\circ}$, i.e. $\simeq 1.9$~\mpcph.

\begin{figure}[ht!]
\centering
\includegraphics[width=0.79\textwidth]{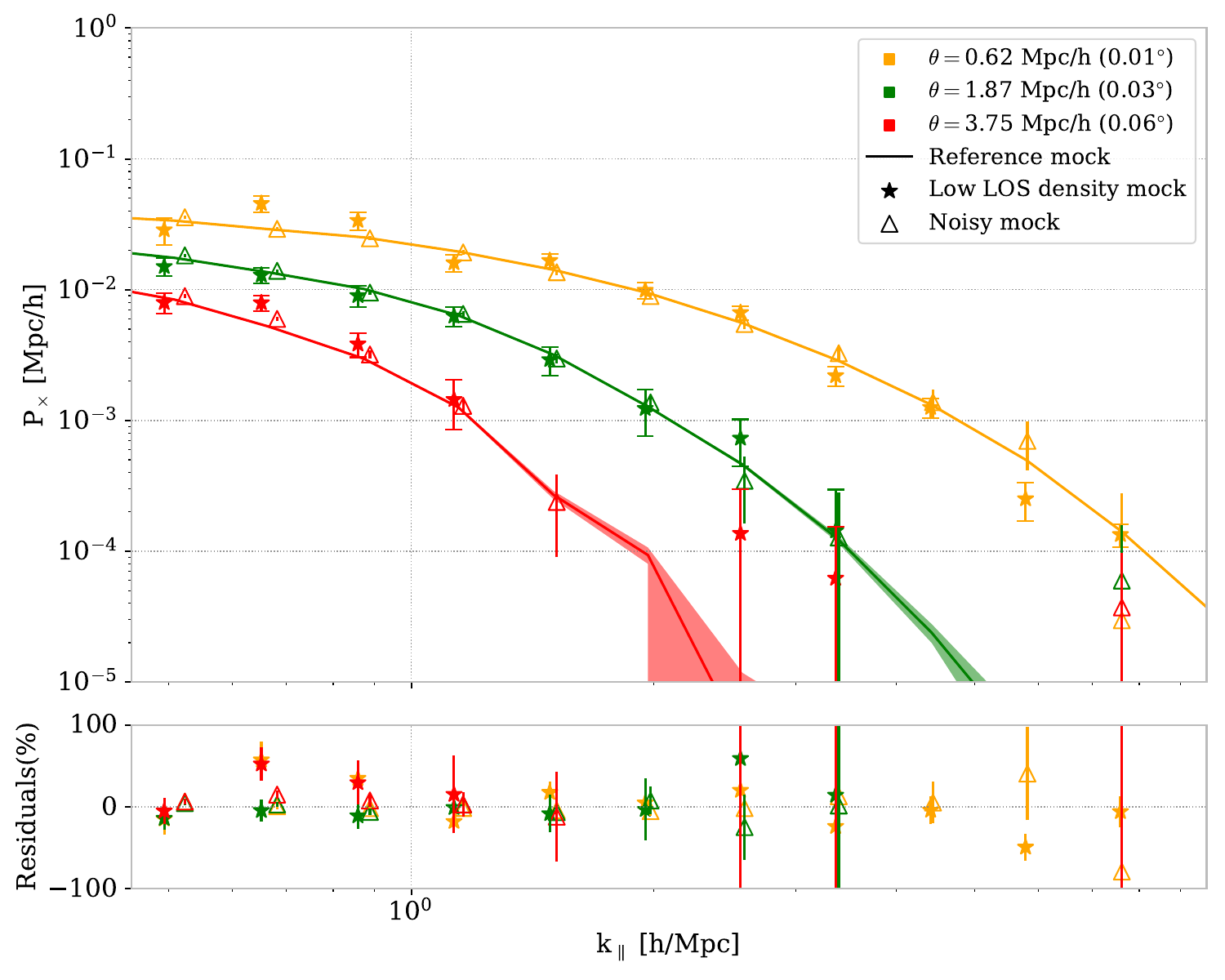}
\caption{Comparison of \px~measurements from different mock realizations, for 3 angular separation bins. The continuous lines represent \px~from the reference mock, already presented in Fig.~\ref{fig:Pcross_10000}, with statistical errors shown by shaded bands. Measurements from mocks with noisy data and with a reduced density of lines-of-sight are shown by triangle points and stars, respectively.  Residuals relative to the reference mock, with their error bars, are shown in the bottom panel. For visual clarity, data points are slightly shifted along the horizontal axis.}
\label{fig:Noise_LOS_combined}
\end{figure}

Figure \ref{fig:Noise_LOS_combined} shows a comparative representation of the estimated \px~from these two mock data sets, in contrast to the reference case of Section~\ref{sec:validation_Nyx}, for three distinct angular separation bins centered at $\theta$ = 0.01$^{\circ}$, 0.03$^{\circ}$ and 0.06$^{\circ}$.
Comparing the \px~from these more realistic mocks to the reference \px, we notice that the considered effects result in an increase of statistical uncertainties in the measurement. More specifically, for the mock sample with noise, the increase of statistical error bars becomes significant when \px~reaches low values: this is similar to the case of \p1 for large $k_{\parallel}$. On the other hand, for the mock with less lines-of-sight, the reduced number of pairs, especially at small angular separations, results in large statistical fluctuations for all measured values.

As a result of these mock studies, we conclude that measurements of \px~for $\theta > 0$ should be feasible with \lya~data from large surveys, but clearly not at the level of precision of existing \p1 measurements.

\section{Application to SDSS quasar spectra}\label{sec:appli-SDSS}

We now move to the application of the method on the largest currently public \lya~forest data set: the SDSS DR16 quasar spectra. We stress that our goal here is to essentially provide a proof-of-principle of our pipeline from a real, already well-studied \lya~forest sample. We are not looking for the most precise measurement possible from this sample, nor do we attempt any interpretation in terms of IGM physics and cosmology.

\subsection{Data sample}\label{sec:Data_sample}
The SDSS~\cite{SDSS:2006srq,Smee:2012wd,SDSS:2011jap,SDSS:2017yll}, BOSS~\cite{BOSS:2012dmf} and subsequent eBOSS~\cite{Dawson:2015wdb} surveys provide the largest publicly available optical spectroscopic data set of extragalactic objects to date. This data set encompasses a large set of \lya~absorption spectra from quasars with redshifts $z>2.1$ and a relatively homogeneous sky distribution spread all over the SDSS footprint, covering approximately 1/4 of the sky.

In our analysis, we select quasars drawn from the eBOSS DR16Q catalog~\cite{Lyke:2020tag}, a component of the 16th SDSS data release~\cite{DR162019}. Our selection criteria are focused on quasars with redshift $z>2.1$, with no identified broad absorption lines (BAL) nor damped Lyman-$\alpha$ (DLA) features in their spectra.

We measure \lya~forest fluctuations, $\delta(\lambda)$, from the set of quasar spectra, using the continuum fitting algorithm implemented in the \texttt{picca}~\footnote{Package for IGM Cosmological-Correlations Analyses, \texttt{https://github.com/igmhub/picca}} code. This code computes $\delta$  from the measured quasar flux $f_q(\lambda)$ through:

\begin{equation}
    \delta(\lambda) = \frac{f_q(\lambda)}{\overline{F}(z) C_q(\lambda)} -1
\end{equation}

\noindent In this expression, $\overline{F}(z)$ is the mean IGM transmission at redshift $z=\lambda/\lambda_{\alpha}-1$, as previously defined. $C_q$ is the quasar's unabsorbed flux, called continuum, and assumed to be a universal function of the rest-frame wavelength $\lambda_{\rm RF}$, multiplied by a quasar-dependent correction: $C_q = (a_q + b_q \log \lambda_{\rm RF})\,C(\lambda_{\rm RF})$. The product $\overline{F}(z)C_q(\lambda)$ is determined by an iterative fit. The continuum fitting procedure is analogous to the approach employed during the eBOSS P1D measurement in~\cite{Chabanier:2018rga}. Specifically, the fitting was done in the rest-frame wavelength range $\lambda_{\rm RF} \in [1050 - 1180]$~\AA. No weights were applied to individual pixel fluxes $f_q(\lambda)$. Two types of pixels were masked: those flagged by the SDSS processing pipeline due to e.g.~cosmic rays, and those that correspond to observed wavelengths matching atmospheric emissions and galactic absorption lines, identically to~\cite{Chabanier:2018rga}. Finally, only spectra with a mean SNR per pixel larger than 1 were used. 

The set of $\delta(\lambda)$ obtained from the DR16 sample is similar to the one used in the P1D measurement of~\cite{Chabanier:2018rga}. Since no resampling is performed at the stage of continuum fitting, all the $\delta(\lambda)$ are pixelized on the SDSS common wavelength grid, equally spaced in logarithmic scale, with a binning $\Delta \log \lambda_{\rm pix} = 10^{-4}$. 

Since our measurement is done separately for different redshift bins as already mentioned in Section~\ref{sec:pcross}, our initial step involves extracting subsets of \lya~forests from the common sample of $\delta$, corresponding to specific redshift bins: for a given redshift $z$, we define a wavelength range $(\lambda_{\rm min} - \lambda_{\rm max})$, centered at $(1+z)\lambda_{\alpha}$ and then, we only select lines-of-sight whose \lya~forest pixels completely overlap this wavelength interval. The width of this interval determines the smallest measurable $k_{\parallel}$. Additionally, lines-of-sight having masked pixels in this wavelength range are discarded. These selections, especially the first one, certainly reduce the statistics. However they greatly simplify the calculation of cross-correlations $\delta_{i} \times \delta_{j}^{*}(k_{\parallel})$ as emphasized in Section~\ref{sec:estimator}.

As in~\cite{Palanque-Delabrouille:2013gaa,Chabanier:2018rga}, given that the spectral binning $\Delta \log \lambda_{\rm pix}$ is in logarithmic scale, we tabulate $k_{\parallel}$ in units of km/s, as already described in Section~\ref{sec:Units_conversions}. The spectroscopic resolution $R$ is derived for each sample using the \texttt{picca} software, identically to~\cite{Palanque-Delabrouille:2013gaa}: it is taken from the spectroscopic pipeline output, and an analytic correction function is applied, depending on the wavelength and position of the spectrum on the CCD, as described in Section~\ref{sec:Noise_resolution}. This correction varies from 1 to 10~\%, and is defined as~\cite{Palanque-Delabrouille:2013gaa}:

\begin{equation}\label{eq:Resolution}
    W(k_{\parallel},R) = \exp \left(-\frac{1}{2}(k_{\parallel}R)^2 \right)\times \frac{\sin(k_{\parallel}\Delta v/2)}{k_{\parallel}\Delta v/2}
\end{equation}

\subsection{Measurement}\label{sec:eBOSS_measurement}

We consider here two redshift bins, for which the statistical sample is particularly abundant. 
\begin{itemize}
    \item The bin $z=2.2$: obtained by selecting $(\lambda_{\rm min}, \lambda_{\rm max}) = (3850, 3930)$~\AA. It includes 38461 \lya~forests covering this wavelength range, each forest containing 89 pixels, spanning over a comoving distance of $\simeq 59$~\mpcph.
    \item The bin $z=2.4$: obtained by selecting $(\lambda_{\rm min}, \lambda_{\rm max}) = (4056, 4210)$~\AA. It includes 18653 \lya~forests covering this wavelength range, each forest containing 162 pixels, spanning over a comoving distance of $\simeq 105$~\mpcph.
\end{itemize}
The choice of wavelength ranges is made with the intention of aligning the data around the chosen redshift values while ensuring the exclusion of spectral regions contaminated by sky lines and calcium H and K galactic absorption lines. Although the sample at $z=2.2$ offers the largest number of lines-of-sight, the $z=2.4$ sample is less affected by noise and contains more pixels per line-of-sight. The width of these spectral segments is sufficiently broad to enable measurements at $k_{\parallel} \gtrsim 0.002$~km/s.

To compute \px, we select $\delta_i$ values based on their mean signal-to-noise per pixel $\overline{\rm SNR}$, to be used in the measurement. For $P_{\times}(\theta > 0)$, we require $\overline{\rm SNR} > 1$. It is worth noting that this criterion is somewhat less stringent compared to the one employed in~\cite{Chabanier:2018rga}. However, our measurement for small, non-zero values of $\theta$, is constrained by limited statistics, and furthermore, noise is expected to have a reduced impact on the cross-correlations between distinct lines-of-sight compared to their auto-correlations. To set this threshold, we scanned over a range of values for the $\overline{\rm SNR}$ threshold and we observed that the average statistical uncertainty in the measured \px~is roughly optimized for the chosen threshold. In the case of $P_{\times}(\theta = 0)$, i.e.~\p1, we have more abundant statistics, and in that instance we employ a more strict criterion: $\overline{\rm SNR} > 4.1$ for $z=2.2$ and $\overline{\rm SNR} > 3.9$ for $z=2.4$.

Additionally, in previous works~\cite{Palanque-Delabrouille:2013gaa,Chabanier:2018rga}, a threshold was applied to each line-of-sight's mean spectroscopic resolution, $\overline{\rm R} > 85$ km/s. We evaluated the impact of such a threshold on our measurements and found it to be small in comparison to our statistical uncertainties for $\theta>0$. Therefore, we apply this threshold only for $P_{\times}(\theta = 0)$ in our study.
Finally, when we compute the \px, we take into consideration both uncorrelated noise and resolution effects, following the procedures detailed in Section~\ref{sec:Noise_resolution}.

\begin{figure}[ht]
\centering
  \includegraphics[width=0.49\textwidth]{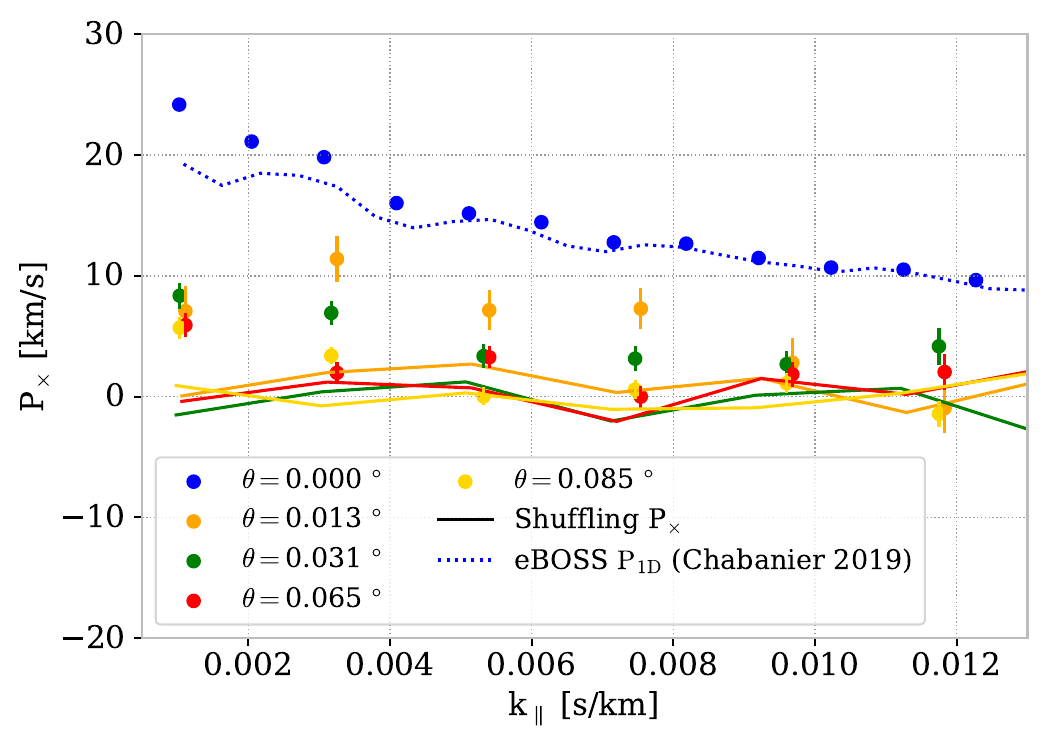}
\centering
\includegraphics[width=0.49\textwidth]{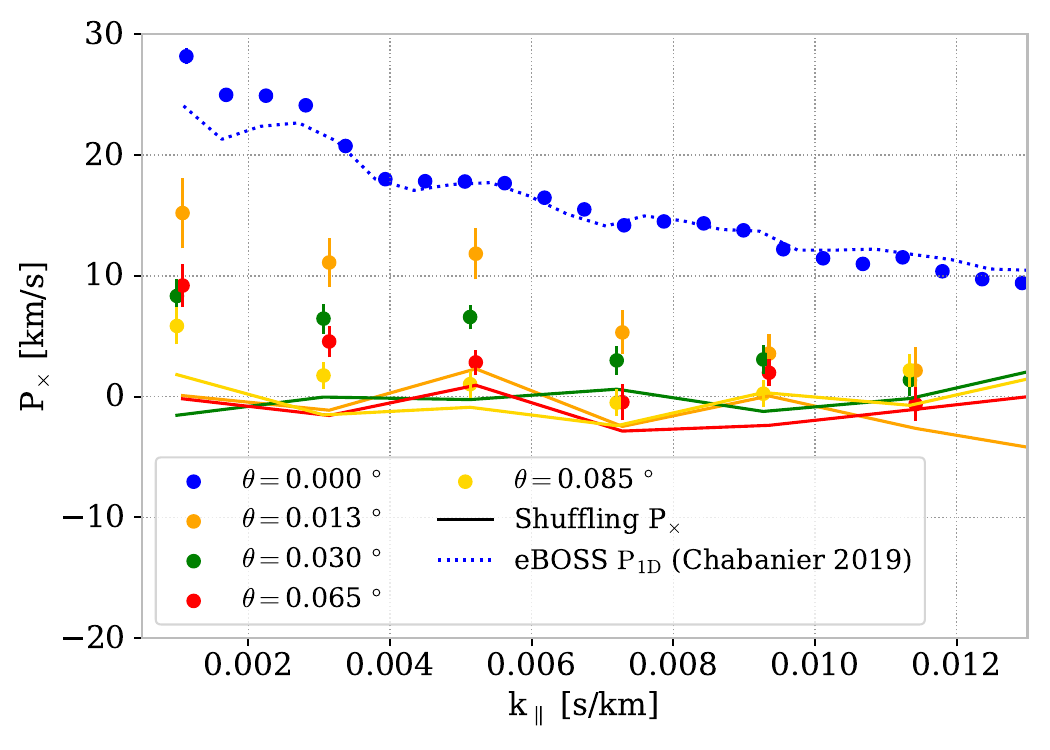}
\caption{\px~measurement from SDSS DR16 data at mean redshifts $z=2.2$ (left) and $z=2.4$ (right). Data points are rebinned as a function of $k_{\parallel}$ for $\theta>0$. Error bars are statistical only. The dotted line represents the published \p1 measurement by~\cite{Chabanier:2018rga}, which includes several additional systematic corrections. Continuous lines represent a null test carried out by shuffling angular coordinates of individual lines-of-sight. Note that the inverse of a pixel size in the original SDSS spectra is $0.0145$ in s/km unit.}
\label{fig:P_cross_eBOSS}
\end{figure}

The cross-spectra measured in the redshift bins $z=2.2$ and 2.4 are given in Fig.~\ref{fig:P_cross_eBOSS}. The error bars represent only statistical uncertainties. The published measurement of \p1 from eBOSS~\cite{Chabanier:2018rga}, depicted in dotted lines, can be compared to our estimation of $P_{\times}(\theta = 0)$. A discrepancy is observed at small values of $k_{\parallel}$, and this is expected due to several factors. To start with, we did not implement a correction due to the continuum fitting. Additionally, we used the SDSS DR16 data set, whereas~\cite{Chabanier:2018rga} relied on the DR14 data set. Furthermore, we employed a different DLA catalog for masking. Lastly, our measurement includes absorption features from the \lya~forest, along with a subdominant contribution from metals that we have not accounted for: metal correlations\footnote{Correlations between \lya~and Si lines are not subtracted, and visible as wiggles in both our measurement and that of~\cite{Chabanier:2018rga}.} were subtracted using sideband data in~\cite{Chabanier:2018rga}, resulting in $\lesssim 10$~\% corrections.

The angular separation bins we choose to compute \px~are centered around 0.013$^{\circ}$, 0.01$^{\circ}$, 0.03$^{\circ}$, followed by a regular binning of width 0.01$^{\circ}$ ranging from 0.05$^{\circ}$ to 0.29$^{\circ}$, and finally 0.35$^{\circ}$, 0.45$^{\circ}$ and 0.55$^{\circ}$. Pair statistics for some of these bins are given in Table~\ref{tab:Stat_eBOSS}. The comparison between the small number of pairs used to compute \px~at small non-zero angular separations and the case where $\theta=0$ (\p1), highlights how challenging this measurement is.

\begin{table}[ht]
\centering
\begin{tabular}{ c|c|cccc } 
 $\langle \theta \rangle (^{\circ})$  & 0  & 0.013  & 0.03  & 0.065  & 0.085  \\ 
 $\langle \theta \rangle$ (\mpcph) & 0  & 0.88  & 2.08  & 4.38  & 5.73  \\ 
 \hline
 $N(z=2.2)$ & 6848 & 105 & 329 & 342 &  440 \\
 $N(z=2.4)$ & 3438 & 28 & 89 & 78 &  104 \\
\end{tabular}
\caption{Number of line-of-sight pairs for some angular separation bins from the selected SDSS data samples, at mean redshifts $z = 2.2$ and $z = 2.4$. The first separation bin $\theta=0$ corresponds to the total number of selected individual lines-of-sight. The conversion of $\theta$ values from degrees to comoving transverse distances in \mpcph~was done using the reference cosmology given in~\ref{sec:description} at $z = 2.3$}.
\label{tab:Stat_eBOSS}
\end{table}

Nevertheless, our analysis reveals that for both $z=2.2$ and 2.4, a clear detection is found across a range of $\theta$ and $k_{\parallel}$ values. To confirm this detection, we perform a null test, which we refer to as the shuffling test. In this test, we permute the angular separations assigned to each line-of-sight pair. Consequently, the shuffled \px~is the result of correlations between lines-of-sight that are effectively separated by large angular distances over the sky. The resulting shuffled \px~values are depicted as solid lines in Fig.~\ref{fig:P_cross_eBOSS}: notably, their average hovers around zero, and it is evident that the distribution of unshuffled \px~points differs from that of the shuffled \px. Moreover, we checked that the measured dispersion of the shuffled \px~points is compatible with our corresponding calculated statistical error bars.

\begin{figure}[ht]
\centering
\includegraphics[width=0.79\textwidth]{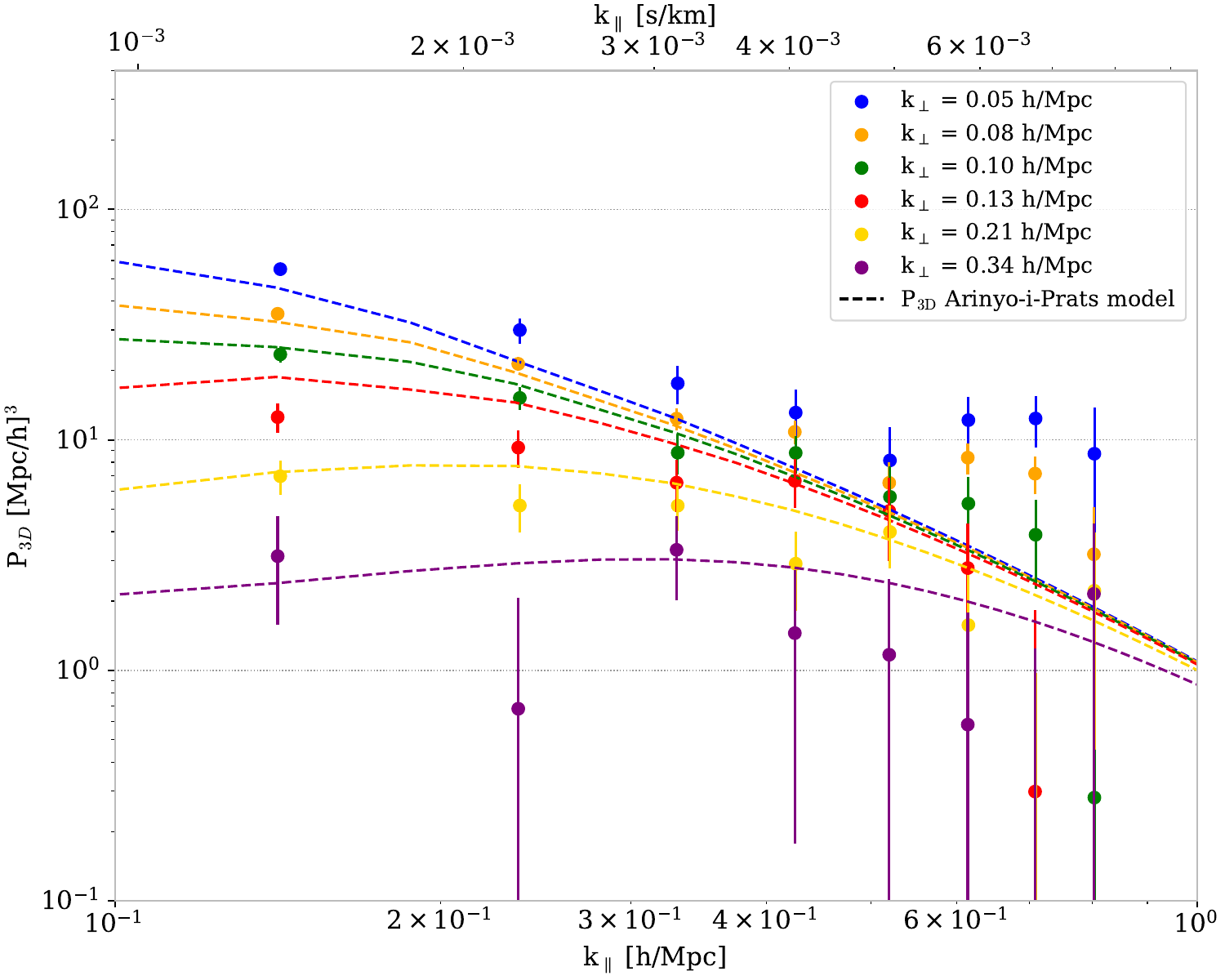}
\caption{Average \p3 measurement from SDSS DR16 data at $z\simeq2.3$, derived from the \px~measurements at $z=2.2$ and $z=2.4$. We converted all units to \mpcph, using the formulae in~\ref{sec:Units_conversions} and the reference cosmology given in~\ref{sec:description}. For illustration, dashed lines represent the \p3 model of Eqn.~\ref{eq:P3D_analytic}, setting $\beta$ = 1.735 and all other parameters as given in Tables 2 and 3 of~\cite{Arinyo-i-Prats:2015vqa} (Fiducial simulation, $z=2.4$).}
\label{fig:P3D_eBOSS_average_2.2_2.4}
\end{figure}

Finally, we infer a \p3 from the statistical combination of the \px~measurements at $z=2.2$ and 2.4. This is done to improve the statistical significance of the measurement. The resulting \p3, shown in Fig.~\ref{fig:P3D_eBOSS_average_2.2_2.4}, therefore corresponds to a measurement at $z\simeq 2.3$.
Error bars in Fig.~\ref{fig:P3D_eBOSS_average_2.2_2.4} represent statistical uncertainties, derived as outlined in Section~\ref{sec:p3d}. In addition, given our method of \p3~estimation, there are important correlations when considering measurement points with identical $k_{\parallel}$ and nearby $k_{\perp}$ values. According to our calculated covariance matrix, the average correlation for pairs of measurements at neighboring $k_{
\perp}$ values in Fig.~\ref{fig:P3D_eBOSS_average_2.2_2.4} is 37~\%. Following the prescription introduced in Section~\ref{sec:p3d}, we compute \p3~for $k_{\parallel}<k_{\parallel,{\rm max}}$, where we find $k_{\parallel,{\rm max}}=0.009$~s/km. This is in line with the observation that, as it can be seen in Fig.~\ref{fig:P_cross_eBOSS}, our measurements of $P_{\times}(\theta>0)$ are statistically compatible with zero for $k_{\parallel} > k_{\parallel,{\rm max}}$.

For an illustrative purpose only, we also represent the \p3~model curves from Eqn.~\ref{eq:P3D_analytic}, with parameters matching those of~\cite{Arinyo-i-Prats:2015vqa} (Tables 2 and 3, Fiducial simulation) at $z=2.4$, with the exception of the $\beta$ parameter, which we set to 1.735. The reasonably close alignment between these curves and the data points shows that our measurement provides physically sound results.

\paragraph{Correlated sky-subtraction noise}\label{sec:PPE_CN}~\\
A detailed study of systematic effects on these \px~and \p3 measurements goes beyond the scope of this article. Statistical fluctuations are certainly the main limiting factor of the measurement at this stage. We expect several of the known systematic effects impacting \p1 to have a similar effect on \px.

On the other hand, let us focus here on an effect which was not impacting previous \p1 measurements: the correlated noise between two spectra contributing to a given line-of-sight pair $(i,j)$. As mentioned in Section~\ref{sec:Noise_resolution}, to compute \px, we assumed $\langle \Tilde{\delta}_{i,n} \Tilde{\delta}_{j,n}^* \rangle = 0$ for $i\neq j$. In the case of SDSS \lya~data, noise correlations between nearby lines-of-sight were studied in details in~\cite{Bautista:2017zgn,duMasdesBourboux:2020pck}. Most of the correlations come from the sky subtraction procedure: all the spectra from a given exposure, located on the same half-plate (spectrograph), are sky-subtracted with a common sky model. This sky model is obtained from a limited sample of sky spectra, whose fluctuations are therefore imprinted on all quasar spectra from the same half-plate. Given the SDSS survey strategy, a large fraction of nearby quasars were observed in the same exposure. This results in artificial correlations in the $\delta$ field for the transverse separations considered in this study. In~\cite{duMasdesBourboux:2020pck}, this was measured on DR16 data and parameterized using the following equation:

\begin{equation}\label{eq:xi_sky}
    \xi_{\rm 3D,sky}(r_{\parallel}, r_{\perp}) = \frac{A_{\rm sky}}{\sigma_{\rm sky} \sqrt{2\pi}} \exp\left( -\frac{r_{\perp}^2}{2\sigma_{\rm sky}^2}\right) \quad\quad {\rm for}\; r_{\parallel}=0, \quad {\rm else}\; 0
\end{equation}

\noindent The parameters are $A_{\rm sky} = 9\times10^{-3}$, $\sigma_{\rm sky} = 31 \,h^{-1}$~Mpc, according to the result of a full fit of $\xi_{\rm 3D}$ in the \lya~forest. Also, this expression is for a binned measurement of $\xi_{\rm 3D}$: the longitudinal binning $\Delta r_{\parallel} = 4\,h^{-1}$~Mpc corresponds to a velocity bin $\Delta v = H(z)\, \Delta r_{\parallel}/(1+z)$. From this expression for $\xi_{\rm 3D,sky}$, we therefore derive:

\begin{equation}\label{eq:px_sky}
P_{\times,{\rm sky}}(r_{\perp},k_{\parallel}) = \frac{\sqrt{2\pi}\Delta v\,A_{\rm sky}}{\sigma_{\rm sky}} \exp\left( -\frac{r_{\perp}^2}{2\sigma_{\rm sky}^2}\right)
\end{equation}

\noindent This is an additive contribution, independent of $k_{\parallel}$. It has the same angular dependence as $\xi_{\rm 3D,sky}$, which is relatively small in our case since $\sigma_{\rm sky} \sim 0.5^{\circ}$, and we consider angular separations much smaller than this value. The order of magnitude is $P_{\times,{\rm sky}}\sim 0.3$~km/s. This is a non-negligible contribution, especially for large $k_{\parallel}$, which should be considered in future measurements. In our case, it remains subdominant with respect to statistical errors. As a cross-check, we performed a data-split test: the angular pair sample was split into same half-plate, and different half-plate pairs. We found no significant difference between the \px~measurements from both sub-samples. This confirms that this measurement is not strongly contaminated by correlated sky-subtraction noise.

\section{Conclusion and outlook}

In this article, we described the implementation of a new approach to measure the small-scale 3D power spectrum of the \lya~forest, \p3. The method relies on using the cross-power spectrum $P_{\times}(\theta, k_{\parallel})$, which can be estimated by correlating 1D Fourier transforms, $\Tilde{\delta}_i(k_{\parallel})$, of individual line-of-sight samples according to Eqn.~\ref{eq:Pcross_3}. This is a straightforward extension of the 1D FFT method widely used to estimate \p1. Its simplicity, both conceptually and in terms of numerical execution, is its key advantage. Compared to the quadratic estimator approach implemented in~\cite{Font-Ribera:2017txs}, our method demands significantly less computational time.

We validated our pipeline using simulated samples of \lya~forests, created from a hydrodynamical simulation box of size 150~\mpcph~at z = 2.0. Although the cosmological volume covered by this simulation sample is small with respect to the SDSS or DESI surveys footprints, it is good enough to measure \p3~at the ``small'' scales considered here. To strengthen our proof-of-principle, we also applied our method to the eBOSS DR16 \lya~forest data, for two redshift bins centered at $z=2.2$ and $z=2.4$. Unlike the case of \p1, the small number of nearby line-of-sight pairs in the data sets, results in very large statistical error bars. Still, we were able to measure for the first time \px~and \p3 for $(k_{\parallel}, k_{\perp})$ values in the $\sim 0.1 - 1\,h~{\rm Mpc^{-1}}$ range. Our measurement fairly compares with analytical formulae derived from simulations.

The statistical limitation of this measurement contrasts with the case of \p1. Large \lya~forest samples from upcoming spectroscopic surveys, like DESI~\cite{DESI:2016fyo} and WEAVE~\cite{WEAVE:2016rxg} will offer good opportunities to measure \px~and \p3: with respect to SDSS, the line-of-sight density of these surveys is expected to increase by a factor of $\sim 4$, so that the pair statistics at small separation angles will be larger by more than an order of magnitude. This will lead to a substantial decrease of the statistical uncertainties.

Clearly, the impact of systematic effects on those measurements should be carefully assessed in future work. However, we expect that these effects will be similar to those affecting \p1~measurements. In some cases, their intensity should naively be smaller than in the case of \p1. For example, the effect of uncorrelated noise is essentially cancelled out when cross-correlating different lines-of-sight. Continuum fitting mostly introduces correlations between $\delta$ pixel values along an individual line-of-sight~\cite{Viel:2001hd}, however it may generate ``distortion'' effects similar to those encountered in measurements of the 3D correlation function~\cite{duMasdesBourboux:2020pck}. The same should hold for high-column density systems~\cite{Rogers:2017eji}. One exception is the impact of correlated noise, since nearby quasars are often observed during the same exposure with large multi-object spectrographs. In this article we presented an order-of-magnitude estimation of this effect in the case of SDSS data.

The physical interpretation of \px~and \p3~measurements is left for future work. There are several possible approaches to fitting these data to models tightly connected to the way \p3~is inferred from \px:
\begin{itemize}
    \item The most straightforward approach will involve directly fitting \px~using a full modelling, which is essentially a simple extension of that used for \p1~alone. This is the most natural strategy since \px~is our primary observable. In particular, the impact of observational systematic effects will be best understood in separate $(\theta,k_{\parallel})$ bins. Additionally, computing \px~from hydrodynamical simulation outputs is as easy as computing \p1 or \p3. This approach is also best suited to understand the differences between contributions of \p1~alone and contributions of transverse correlations, to physical constraints on IGM and cosmological parameters.
    \item The method used in this article to estimate \p3~from \px~is a simple and qualitative approach which is not statistically optimal. It will be possible to infer \p3~from \px~using a full likelihood, or quadratic estimator inference, identically to what was pioneered in~\cite{Font-Ribera:2017txs}.
    \item Finally, using either \px~or \p3, it should be possible to fit analytic formulae that connect \p3~to the linear matter power spectrum such as proposed in~\cite{McDonald:2001fe,Arinyo-i-Prats:2015vqa,Garny:2020rom}. Correlations between fit parameters should be reduced with respect to the case of \p1 alone, thanks to the $(k_{\parallel},k_{\perp})$ dependence of \p3. This should provide a fast, but hopefully reasonably accurate way to infer the properties of the linear power spectrum without resorting to a full simulation-based approach.
\end{itemize}

\noindent The interpretation of \px~or \p3 measurements is expected to disentangle various physical effects that are at play in the \lya~forest, and that are too degenerate to be robustly separated using \p1~alone~\cite{Rorai:2013dxa}. This is in particular the case of degeneracies between a possible primordial cut off in the power spectrum, associated for example to a Warm Dark Matter scenario, and thermal properties of the IGM~\cite{Garzilli:2015iwa}. We anticipate that, using our method, precise measurements of \px~and \p3 are within reach with the upcoming spectroscopic data from DESI and WEAVE.

\acknowledgments
We thank Jim Rich for engaging in numerous insightful discussions and providing valuable feedback at both the initial and advanced phases of this project. We thank Andreu Font Ribera and Christophe Yèche for their constructive discussions. SC and ZL are partially supported by the DOE’s Office of Advanced Scientific Computing Research and Office of High Energy Physics through the Scientific Discovery through Advanced Computing (SciDAC) program. EA acknowledges support from ANR/DFG grant ANR-22-CE92-0037 for the DESI-Lya project. CR acknowledges support from Excellence Initiative of Aix-Marseille University - A*MIDEX, a French "Investissements d'Avenir" program (AMX-20-CE-02 - DARKUNI). This project makes use of the \texttt{picca} software~\cite{duMasdesBourboux:2020pck}. This research used resources of the National Energy Research Scientific Computing Center, a DOE Office of Science User Facility supported by the Office of Science of the U.S. Department of Energy under Contract No. DEC02-05CH11231.

\bibliographystyle{JHEP.bst}
\bibliography{biblio} 

\end{document}